\documentclass[showkeys,notitlepage,superscriptaddress,floatfix,prd,10pt,aps]{revtex4-1}

\usepackage[utf8]{inputenc}
\usepackage[colorlinks=true,citecolor=blue,linkcolor=blue]{hyperref}
\usepackage[normalem]{ulem}
\usepackage{url}
\usepackage{graphicx,wrapfig,float,slashed,cancel}
\usepackage{amsmath,amssymb,epsfig,graphicx,xcolor,stmaryrd}
\usepackage{bbold} 
\usepackage{bm}
\usepackage{enumitem}
\usepackage{multirow}

\usepackage{physics}
\usepackage{orcidlink}

\definecolor{darkblue}{RGB}{1, 90, 173}

\def\pslash{p\!\!\!\slash }

\def\qslash{q\!\!\!\slash }
\def\xslash{x\!\!\!\slash }
\def\yslash{y\!\!\!\slash }
\def\dslash{D\!\!\!\slash }

\begin{document}
	
	
\title{Mechanical properties of the $\Omega^-$ baryon from gravitational form factors}
\author{Zeinab Dehghan\orcidlink{0000-0002-5976-9231}}
\email{zeinab.dehghan@ut.ac.ir}	\thanks{Corresponding author}
\affiliation{Department of Physics, University of Tehran, North Karegar Avenue, Tehran 14395-547, Iran}
\affiliation{School of Physics, Institute for Research in Fundamental Sciences (IPM) P.O. Box 19395-5531, Tehran, Iran}
\author{K. Azizi\orcidlink{0000-0003-3741-2167}}
\email{kazem.azizi@ut.ac.ir}
\thanks{Corresponding author}
\affiliation{Department of Physics, University of Tehran, North Karegar Avenue, Tehran
14395-547, Iran}
\affiliation{Department of Physics, Dogus University, Dudullu-\"{U}mraniye, 34775
Istanbul,  T\"{u}rkiye}
\affiliation{School of Particles and Accelerators, Institute for Research in Fundamental Sciences (IPM), P.O. Box 19395-5746, Tehran, Iran}

\date{\today}
	
\preprint{}
	
\begin{abstract}

We present a comprehensive investigation of the mechanical properties of the $\Omega^-$ baryon by analyzing its gravitational form factors (GFFs) within the framework of QCD sum rules. 
These form factors encode rich information about the internal structure of hadrons and offer deep insights into the dynamics that govern their stability. 
The spin‑3/2 nature of the $\Omega^-$ baryon manifests in its gravitational form factors as intricate multipole structures, which encapsulate higher‑order deformations and demonstrate the influence of intrinsic spin on internal dynamics.
We extract the GFFs of the $\Omega^-$ baryon and apply their specific multipole combinations, gravitational multipole form factors (GMFFs), to quantify key mechanical observables—including energy density, angular momentum, pressure and shear force distributions, mass and mechanical radii, and D-terms—associated with different multipole orders. 
Notably, this work provides the first determination of several of these observables, such as the mechanical radii and the quadrupole contributions to the pressure and shear force distributions.
Our analysis shows that the quadrupole contributions to the mechanical properties are generally subdominant compared to those from the monopole component.
We further investigate the mechanical stability of the $\Omega^-$ through a multipole analysis of its internal force distributions.
These results enhance our understanding of the mechanical structure of spin-3/2 hadrons and provide useful benchmarks for future theoretical and lattice QCD studies.

\end{abstract}
	
	
	
\maketitle
	
\renewcommand{\thefootnote}{\#\arabic{footnote}}
\setcounter{footnote}{0}
\section{Introduction}\label{intro} 
	
Probing the structure of hadrons remains a major challenge in quantum chromodynamics (QCD) and is essential for uncovering the role of quark–gluon dynamics in strong interactions. 
While electromagnetic form factors have traditionally served as primary tools for revealing the charge and magnetization distributions of quarks inside hadrons, they are inherently limited in providing a complete internal picture, particularly regarding the mechanical aspects of hadronic structure. 
A detailed investigation of the mechanical properties of hadrons offers crucial insight into the distribution of internal forces and the dynamics that govern their stability and shape their geometric configurations.
Gravitational form factors (GFFs) are fundamental quantities that parameterize the matrix elements of the energy-momentum tensor (EMT) between hadronic states. 
These form factors yield a deeper understanding of the mechanical structure of hadrons by granting access to the spatial distributions of energy, angular momentum, pressure, and shear forces within them~\cite{Kobzarev:1962wt, Pagels:1966zza, Polyakov:2002yz, Polyakov:2018zvc, Burkert:2023wzr}.
By capturing the balance of forces that confine constituents, GFFs provide valuable information about the non-perturbative dynamics of strong interactions.

Gravitational form factors are currently inaccessible to direct experimental measurement due to the extremely weak nature of gravitational interactions at the hadronic scale. Nevertheless, these form factors can be explored indirectly through their established connection to Generalized Parton Distributions (GPDs), which encode rich information about the internal structure of hadrons
~\cite{Collins:1996fb, Ji:1996ek, Ji:1996nm, Radyushkin:1997ki, Anikin:2017fwu, Burkert:2023wzr, Kumericki:2019ddg, Burkert:2018bqq, Selyugin:2023hqu, Wang:2024sqg, Dutrieux:2021nlz, Wang:2024fjt, Goharipour:2025lep, Goharipour:2024mbk, Hashamipour:2022noy, Mamo:2019mka, Bhattacharya:2023ays, Won:2023ial, Shuryak:2023siq, Chen:2024adg, Bhattacharya:2023wvy, Ji:2021mfb, Freese:2021czn, Freese:2021qtb, Burkert:2021ith, Son:2024uet, Mamedov:2024tth, Tandy:2025tea, Fu:2022bpf, Fu:2023dea, Fu:2024kfx, Kumano:2017lhr}. GPDs are studied through high-energy exclusive processes such as Deeply Virtual Compton Scattering (DVCS) and Deeply Virtual Meson Production (DVMP), offering an indirect yet powerful means to extract the moments that determine GFFs~\cite{Burkert:2021ith, Collins:1996fb, Ahmed:2024grm}.  
Beyond experimental extractions, GFFs are actively investigated within a variety of theoretical frameworks that capture the non-perturbative features of quark-gluon interactions in various hadronic systems, including lattice QCD, QCD sum rules, chiral effective field theory (EFT), the Skyrme model, light-front quark-diquark models, and other phenomenological approaches
~\cite{Bali:2018zgl,Alexandrou:2019ali,Lin:2023ass, GarciaMartin-Caro:2023klo, Kou:2023azd, Alharazin:2023uhr, GarciaMartin-Caro:2023toa, Yao:2024ixu, Cao:2023ohj, Panteleeva:2022uii, Panteleeva:2021iip, Panteleeva:2020ejw, Won:2022cyy, Hatta:2023fqc, Choudhary:2022den, Neubelt:2019sou, Tong:2022zax, Cao:2024fto,Hagiwara:2024wqz, Nair:2025sfr, Sugimoto:2025btn, Sain:2025kup, Fujii:2025aip, Cao:2025dkv, Kaur:2025gyr, Pentchev:2025qyn, Fujii:2024rqd, Sun:2025ank, Hatta:2025ryj, Tanaka:2018wea}.

The energy-momentum tensor in QCD naturally decomposes into quark and gluon parts~\cite{Freese:2021jqs, Freese:2025glz}, each contributing uniquely to the internal dynamics and mechanical features of hadrons. In principle, a complete and consistent description of hadron structure requires the consideration of both quark and gluon contributions, as the total EMT current obeys the conservation law, expressed by ${\partial}^{\mu} T_{\mu\nu} = 0$. This full picture is crucial for understanding the interplay between quarks and gluons in generating the observed mechanical stability and internal force balance inside hadrons. Moreover, studying quark and gluon contributions separately provides valuable insights into how each component influences the mechanical properties. 
Combining full quark-gluon studies with individual component investigations offers a comprehensive view of non-perturbative QCD dynamics and the internal stability of hadrons. 
Significant efforts have been devoted to evaluating the gluon contributions to the energy-momentum tensor current~\cite{Meziani:2024cke, More:2023pcy, Guo:2023pqw, Guo:2023qgu, Duran:2022xag, Fan:2022qve, Mamo:2021krl, Mamo:2019mka, Sain:2025kup, Tong:2021ctu, Shanahan:2018pib, Pefkou:2021fni, Pentchev:2025qyn, Tandy:2025tea} and the quark contributions~\cite{Burkert:2021ith, Lorce:2023zzg, Anikin:2019kwi, Azizi:2019ytx, Dehghan:2025ncw, Tanaka:2018wea, Freese:2019bhb, Aliev:2020aih} within various hadronic systems and theoretical approaches. Moreover, the quark and gluon components have been examined both individually~\cite{More:2023vlb, Wang:2023fmx} and in combination~\cite{Selyugin:2023hqu, Hackett:2023rif, Yao:2024ixu}.

Gravitational form factors have been defined and studied for hadronic systems with different spin, providing essential insight into how spin influences the internal mechanical properties of hadrons. 
For spin-0 hadrons such as the pion and kaon~\cite{Tanaka:2018wea, Kumano:2017lhr, Freese:2019bhb, Fujii:2024rqd, Hatta:2025ryj, Aliev:2020aih, Shanahan:2018pib, Pefkou:2021fni, Kaur:2025gyr, Cao:2025dkv, Voronin:2025sbs, Fujii:2025tpk}, spin-1/2 systems like the nucleon~\cite{Ji:1996ek, Bali:2018zgl, Neubelt:2019sou,Alexandrou:2019ali, Panteleeva:2020ejw, Burkert:2021ith, Panteleeva:2021iip, Won:2022cyy, Panteleeva:2022uii, Choudhary:2022den, Tong:2022zax,Lin:2023ass, GarciaMartin-Caro:2023klo, Kou:2023azd, Alharazin:2023uhr, GarciaMartin-Caro:2023toa, Cao:2023ohj, Hatta:2023fqc, Hackett:2023rif, Lorce:2023zzg, Yao:2024ixu, Cao:2024fto, Hagiwara:2024wqz, Nair:2025sfr, Sugimoto:2025btn, Sain:2025kup, Fujii:2025aip, Cao:2024zlf, Tanaka:2025pny, Fujii:2025paw}, and spin-1 hadrons including the deuteron and vector mesons such as the $\rho$ meson~\cite{Cosyn:2019aio, Freese:2019bhb, Polyakov:2019lbq, Aliev:2020aih, Sun:2020wfo, Epelbaum:2021ahi, Pefkou:2021fni, Kim:2022wkc, Freese:2022yur, Freese:2022ibw, Mamedov:2024tth}, 
the EMT structure becomes progressively more complex due to the increasing spin and polarization degrees of freedom, which manifests in the growing number of independent gravitational form factors characterizing each system. 
For spin-3/2 hadrons, the energy-momentum tensor exhibits a more intricate decomposition, giving rise to multipole gravitational form factors—$\varepsilon_{0,2}(t), \mathcal{J}_{1,3}(t)$, and $D_{0,2,3}(t)$—which describe the spatial distribution of energy, angular momentum and internal forces, respectively, beyond the monopole level. These form factors encode higher-order deformations and capture the influence of intrinsic spin on the internal dynamics of the system. 
The $D_n(t)$ form factors define the multipole components of the pressure and shear force distributions, denoted by $p_n(r)$ and $s_n(r)$, as well as the associated D-terms $\mathcal{D}_n$ for $n = 0, 2, 3$. These contributions provide valuable insight into the mechanical structure of spin-3/2 baryons, with the D-terms playing a pivotal role in maintaining internal force balance and ensuring mechanical stability.
In addition, the mass and mechanical radii are fundamental quantities that characterize the spatial distributions of energy and internal forces, respectively. In spin-3/2 hadrons, their multipole extensions further reveal the geometric shape and spin-induced deformations of the internal structure.
Although electromagnetic form factors of spin-3/2 states have been extensively studied~\cite{Nozawa:1990gt, Kotulla:2002cg, Pascalutsa:2006up, Azizi:2009egn, Boinepalli:2009sq, Alexandrou:2010jv, Kim:2019gka, Ramalho:2020laj, Fu:2022rkn, Hong:2023tkv, Fu:2025vkq}, investigations of their gravitational form factors remain relatively limited, primarily due to the challenges in experimental access through GPDs, which arises from the short lifetimes of spin-3/2 baryons.
Gravitational form factors and the associated mechanical properties of spin-3/2 hadrons, as well as those relevant to transition processes such as $N \to \Delta$, have been investigated using various non-perturbative methods, including the Skyrme model, lattice QCD, chiral EFT, QCD sum rules, and other phenomenological approaches~\cite{Perevalova:2016dln, Cotogno:2019vjb, Panteleeva:2020ejw, Pefkou:2021fni, Kim:2021zbz, Kim:2020lrs, Fu:2022rkn, Alharazin:2022wjj, Alharazin:2022xvp, Fu:2023ijy, Fu:2023dea, Wang:2023bjp, Dehghan:2023ytx, Kim:2022bwn, Ozdem:2022zig, Kim:2023yhp, Alharazin:2023zzc, Goharipour:2024atx, Sun:2025ank}. In addition, recent progress has been made in the formulation of generalized parton distributions for spin-3/2 systems~\cite{Fu:2022bpf, Fu:2023dea, Fu:2024kfx}, paving the way for future studies of their GFFs within GPD-based frameworks.

Studies on the gravitational form factors of the $\Omega^-$ baryon remain limited. Recently, these form factors have been investigated within a covariant quark-diquark framework~\cite{Fu:2023ijy, Wang:2023bjp}. 
Further theoretical studies are crucial to elucidate the spin-dependent mechanical structure of the $\Omega^-$, a uniquely stable spin-3/2 baryon composed entirely of strange quarks and unaffected by strong decays. 
In this work, we systematically determine the gravitational form factors of the $\Omega^-$ baryon—incorporating both quark and gluon contributions to the energy-momentum tensor—within the QCD sum rules framework, and provide a comprehensive characterization of its mechanical properties.
The QCD sum rule approach is a powerful and well-established nonperturbative method widely employed in hadron physics, especially effective in extracting form factors, coupling constants, and other hadronic parameters.
In the Breit frame, which allows a commonly used three-dimensional spatial interpretation of the energy-momentum tensor, we derive the gravitational multipole form factors (GMFFs) as specific combinations of the GFFs. 
However, it should be emphasized that, in a relativistic system, the hadron’s wave packet cannot be completely factorized from its internal structure, and thus such spatial interpretations have limitations.
More rigorous separation is only achievable within the light-front (LF) formalism, or through phase-space approaches such as Wigner distributions~\cite{Freese:2021czn,Lorce:2020onh}. 
These GMFFs facilitate a systematic multipole decomposition of mechanical observables, through which we extract the multipole structure of the spatial distributions of energy density, angular momentum, pressure, and shear forces, as well as derive the mass and mechanical radii and D-terms.
In particular, by examining the pressure, shear force distributions, and D-terms across different multipole orders, we gain insight into the internal force equilibrium and assess the mechanical stability of the $\Omega^-$ baryon.

This paper is organized as follows. In Sec.~\ref{sec:general}, we introduce the general theoretical framework for the gravitational form factors of spin-3/2 systems and identify the associated physical observables that characterize their internal structure.
In Sec.~\ref{sec:formalism}, we present the calculation of the gravitational form factors of the $\Omega^-$ baryon within the three-point QCD sum rules framework and a detailed numerical analysis of these form factors
is conducted in Sec.~\ref{sec:numerical}.
In Sec.~\ref{sec:mech}, we derive key mechanical observables from the gravitational form factors and explore the internal dynamics governing the stability of the $\Omega^-$ baryon.
Finally, in Sec.~\ref{sec:con}, we conclude by discussing the main findings.

\section{GFFs of spin-3/2 particles }\label{sec:general}
	
In this work, we aim to determine the gravitational form factors (GFFs) of the $\Omega^{-}$ baryon and investigate its mechanical properties using the QCD sum rules approach. To establish the theoretical foundation for this analysis, in this section we present the general formalism of GFFs for spin-3/2 particles and the corresponding physical observables they describe. 
These form factors are defined via the matrix elements of the energy-momentum tensor (EMT), which is decomposed into quark and gluon contributions as,
\begin{equation}
		\label{eq:totalEMT}
		T_{\mu\nu} (z) = \sum_q T_{\mu\nu}^q (z) + T_{\mu\nu}^g (z) ,
\end{equation}
where, for the symmetric (Belinfante) EMT used in our analysis, the individual components are given by~\cite{Polyakov:2018zvc, Freese:2021jqs}:
\begin{eqnarray}
		\label{eq:EMTcurrentquark}
		T_{\mu\nu}^q (z) &=& \frac{i}{2} \bar{q} (z) \bigg(\overleftrightarrow{D}_\mu \gamma_\nu
		+ \overleftrightarrow{D}_\nu \gamma_\mu \bigg) q(z)
		- g_{\mu\nu} \bar{q}(z)  \big(i \overleftrightarrow{\dslash} - m_q\big)q(z),
		\\
		\label{eq:EMTcurrentgluon}
		T_{\mu\nu}^g (z) &=& G_{\mu\rho}(z) G^{\rho}_{\hspace{0.5mm},\nu}(x) + \frac{1}{4}g_{\mu\nu}G^{\rho\delta}(z)G_{\rho\delta}(z).
\end{eqnarray}
Recently, a corrected form of the gauge-invariant, non-symmetrized energy-momentum tensor—derived directly from Noether’s second theorem—is developed in Ref.~\cite{Freese:2025glz}.
The covariant derivative $\overleftrightarrow{D}_{\mu}$ in Eq.~\eqref{eq:EMTcurrentquark} is defined as $\overleftrightarrow{D}_\mu (z) = [\overrightarrow{D}_\mu (z) - \overleftarrow{D}_\mu (z)]/2$ where,
\begin{equation}
		\label{eq:derivdetail}
		\overrightarrow{D}_{\mu}(z)=\overrightarrow{\partial}_{\mu}(z)-i
		\frac{g}{2}\lambda^a A^a_\mu(z), \qquad
		\overleftarrow{D}_{\mu}(z)=\overleftarrow{\partial}_{\mu}(z)+
		i\frac{g}{2}\lambda^a A^a_\mu(z),
\end{equation}
with $A^a_\mu(z)$ denoting the external gluon fields and 
$\lambda^a$ being the Gell-Mann matrices of SU(3). The most comprehensive parametrization of the matrix element of the energy-momentum tensor between spin-3/2 hadronic states involves ten independent gravitational form factors, $F_{i,k}$, and is expressed as~\cite{Cotogno:2019vjb, Kim:2020lrs}:
\begin{align}\label{eq:matrix element}
\begin{aligned}
			\langle {p',s'}
			|T_{\mu \nu}(0)|
			{p,s}\rangle &= 
			- \bar{u}_{\alpha'}(p',s')\Big\{
			\frac{P_{\mu}P_{\nu}}{m}
			\Big(g^{\alpha' \beta'} F_{1,0}(t) 
			- \frac{\Delta^{\alpha'}\Delta^{\beta'}}{2 m^2} 
			F_{1,1}(t)\Big)
			\\
			&\hspace{-2.2 cm}+
			\frac{(\Delta_{\mu}\Delta_{\nu} - g_{\mu\nu} \Delta^2)}{4 m}
			\Big(g^{\alpha' \beta'} F_{2,0}(t) 
			- \frac{\Delta^{\alpha'}\Delta^{\beta'}}{2 m^2} 
			F_{2,1}(t)\Big)
			+ m g_{\mu\nu}
			\Big(g^{\alpha' \beta'} F_{3,0}(t) 
			- \frac{\Delta^{\alpha'}\Delta^{\beta'}}{2 m^2} 
			F_{3,1}(t)\Big)\\
			&\hspace{-2.2 cm}+
			\frac{i}{2}
			\frac{(P_{\mu} \sigma_{\nu\rho} + P_{\nu} \sigma_{\mu\rho})
				\Delta^{\rho}}{m}
			\Big(g^{\alpha' \beta'} F_{4,0}(t) 
			- \frac{\Delta^{\alpha'}\Delta^{\beta'}}{2 m^2} 
			F_{4,1}(t)\Big)
			- \frac{1}{m}\Big(
			g^{\alpha'}_{\mu}\Delta_{\nu}\Delta^{\beta'} 
			+ g^{\alpha'}_{\nu}\Delta_{\mu}\Delta^{\beta'}
			+ g^{\beta'}_{\mu}\Delta_{\nu}\Delta^{\alpha'}
			\\
			&\hspace{-2.2 cm}
			+ g^{\beta'}_{\nu}\Delta_{\mu}\Delta^{\alpha'}
			-2 g_{\mu\nu} \Delta^{\alpha'}\Delta^{\beta'} 
			- \Delta^2 g^{\alpha'}_{\mu} g^{\beta'}_{\nu}
			- \Delta^2 g^{\alpha'}_{\nu} g^{\beta'}_{\mu}
			\Big)F_{5,0}(t)
			+
			m \Big(
			g^{\alpha'}_{\mu} g^{\beta'}_{\nu}
			+ g^{\alpha'}_{\nu} g^{\beta'}_{\mu}
			\Big)F_{6,0}(t)
			\Big\} u_{\beta'}(p,s), 
\end{aligned}
\end{align}
where $u_{\beta'}(p,s)$ denotes the Rarita-Schwinger spinor for a particle carrying momentum $p$ and spin $s$, $P = (p + p')/2$ represents the average momentum, $\Delta = p' - p$ is the momentum transfer, with the invariant momentum transfer squared given by $t =  \Delta^2$, and $m$ is the mass of the particle. Among these form factors, $F_{i,k}$ with $i=1, 2, 4, 5$ correspond to conserved EMT current, while $F_{i,k}$ with $i=3, 6$ are associated with non-conserved contributions. Since we consider the full EMT, which includes both quark and gluon parts, the non-conserved form factors vanish due to EMT conservation, ${\partial}^{\mu} T_{\mu\nu} = 0$, and are therefore omitted from our analysis.
	
To reveal the spatial structure and physical information associated with the gravitational form factors, it is convenient to evaluate the matrix elements of the energy-momentum tensor in the Breit frame. In this frame, the average momentum is purely timelike, while the momentum transfer is purely spatial: 
\begin{equation}\label{eq:Breit}
		P^{\mu} = (E, \vec{0}), \qquad \Delta^{\mu} = (0, \boldsymbol{\Delta}),
		\qquad t = \Delta^2 = - {\boldsymbol{\Delta}}^2 = 4 (m^2 - E^2).
\end{equation}
This choice simplifies the interpretation of the form factors in terms of static mechanical properties of the hadron and allows a clear separation between different components of the EMT. Distinct components of the EMT correspond to specific physical quantities characterizing the internal dynamics of the particle. The temporal component, $T_{00}$, is associated with the energy (or mass) distribution inside the hadron. The mixed components, $T_{0i}$, are related to the spatial distribution of momentum, while the spatial components, $T_{ij}$, encode the internal stress tensor, which describes the distribution of forces such as pressure and shear stress acting within the hadron.
	
For a spin-3/2 particle, the gravitational form factors exhibit a richer structure compared to spin-1/2 systems, incorporating higher multipole contributions such as quadrupole and octupole. These multipoles reflect the particle’s intrinsic spin and reveal how spin influences the internal distributions of energy, momentum, and stress. In the Breit frame, this structure becomes manifest, as the matrix elements of the EMT can be reorganized in terms of gravitational multipole form factors (GMFFs)~\cite{Kim:2020lrs},
\begin{eqnarray}
		\label{eq:matrix element00}
		\langle {p',s'}
		|T^{00}(0)|
		{p,s}\rangle &=& 2mE \Big[\varepsilon_0(t) \delta_{s' s}
		+ {\Big(\frac{\sqrt{-t}}{m} \Big)}^2
		Y^{kl}_2 \hat{Q}^{kl}_{s' s}  \varepsilon_2(t) 
		\Big], \\
		\label{eq:matrix element0i}
		\langle {p',s'}
		|T^{0i}(0)|
		{p,s}\rangle &=& 2mE \Big[
		\frac{\sqrt{-t}}{m}
		i \epsilon^{ikl} Y^l_1 \hat{S}^k_{s' s} \mathcal{J}_1(t)
		+ {\Big(\frac{\sqrt{-t}}{m} \Big)}^3
		i \epsilon^{ikl} Y^{lmn}_3 \hat{O}^{kmn}_{s' s} \mathcal{J}_3(t) 
		\Big],\\
		\label{eq:matrix elementij}
		\langle {p',s'}
		|T^{ij}(0)|
		{p,s}\rangle &=& 2mE \Big[
		\frac{1}{4 m^2} (\Delta^{i}\Delta^{j} + \delta^{ij} \Delta^2) D_0(t) \delta_{s' s}
		+ \frac{1}{4 m^4}
		\hat{Q}^{kl}_{s' s}
		(\Delta^{i}\Delta^{j} + \delta^{ij} \Delta^2) 
		\Delta^{k} \Delta^{l} D_3(t) \nonumber\\
		&+&
		\frac{1}{2 m^2}
		(\hat{Q}^{ik}_{s' s}
		\Delta^{j}\Delta^{k}  
		+\hat{Q}^{jk}_{s' s}
		\Delta^{i}\Delta^{k} 
		+
		\hat{Q}^{ij}_{s' s} \Delta^2 
		-
		\delta^{ij}
		\hat{Q}^{kl}_{s' s}
		\Delta^{k} \Delta^{l} ) D_2(t)
		\Big],
\end{eqnarray}
where $\varepsilon_0(t)$ and $\varepsilon_2(t)$ denote the energy monopole and quadrupole form factors, respectively, while $\mathcal{J}_1(t)$ and $\mathcal{J}_3(t)$ correspond to the angular momentum dipole and octupole form factors. The form factors $D_0(t)$, $D_2(t)$, and $D_3(t)$ encode the information associated with the multipole contributions to the internal pressure and shear force distributions of the spin-3/2 particle. The explicit definitions of these GMFFs, as well as the $n$-rank irreducible tensors $Y^{i_1 \dots i_n}_n$ and the spin-dependent operators $\hat{Q}^{ij}$ (quadrupole) and $\hat{O}^{ijk}$ (octupole) expressed in terms of the spin operator $\hat{S}^i$, are provided in Appendix~\ref{sec:AppGMFFs}.
	
In the Breit frame, the Fourier transform of gravitational form factors provides a powerful tool to access the spatial distributions of energy density, pressure, and shear forces inside hadrons under the static approximation, closely analogous to the electromagnetic case~\cite{ Sachs:1962zzc, Polyakov:2002yz, Polyakov:2018zvc, Miller:2007uy, Freese:2021czn}. 
Both the static energy-momentum tensor densities and the gravitational form factors in coordinate space are obtained through the three-dimensional Fourier transformation of their respective momentum-space expressions~\cite{Polyakov:2018zvc, Kim:2020lrs},
\begin{eqnarray}
\label{eq:EMTspace}
		T^{\mu \nu}(\boldsymbol{r}, s', s) &=& \int \frac{d^3 \boldsymbol{\Delta}}{(2 \pi)^3 2 E} e^{-i \boldsymbol{\Delta}.\boldsymbol{r}} \langle {p',s'}
		|T^{\mu \nu}(0)|
		{p,s}\rangle,\\
		\label{eq:fourier}
		[f(t)]_{FT} &=& \int \frac{d^3 \boldsymbol{\Delta}}{(2 \pi)^3} e^{-i \boldsymbol{\Delta}.\boldsymbol{r}} f(t)
		= \frac{1}{4 \pi^2} \int_{-\infty}^{0} \frac{\sin[r\sqrt{-t}]}{r} f(t) dt,
\end{eqnarray}
where the spherical symmetry of the system simplifies Eq.~\eqref{eq:fourier} to its final form.
In the following, we investigate the spatial distributions of spin-3/2 particles in the Breit frame, taking into account the emergence of higher multipole contributions arising from their intrinsic spin.

The temporal component of the energy-momentum tensor encodes the spatial distributions of the energy density through monopole, $\varepsilon_0(r)$, and quadrupole, $\varepsilon_2(r)$, contributions~\cite{Kim:2020lrs}:
\begin{equation}
\label{eq:EMTspace00}
		T^{00}(\boldsymbol{r}, s', s) = 
		\varepsilon_0 (r) \delta_{s' s}
		+
		\varepsilon_2 (r)
		Y^{kl}_2 (\Omega_r) \hat{Q}^{kl}_{s' s},
\end{equation}
where the radial functions $\varepsilon_0(r)$ and $\varepsilon_2(r)$ are defined as follows,
\begin{eqnarray}
\label{eq:energyr0}
	\varepsilon_0 (r) &=& m [\varepsilon_0 (t)]_{FT},\\
\label{eq:energyr2}
	\varepsilon_2 (r) &=& -\frac{1}{m} r \frac{d}{dr} \frac{1}{r} \frac{d}{dr} [\varepsilon_2 (t)]_{FT}.
\end{eqnarray}
The monopole component of the energy density exhibits spherical symmetry, whereas the quadrupole component characterizes the deformation of the energy distribution from the spherically symmetric configuration. For a particle with arbitrary spin, the general tensor quantity can be defined~\cite{Cosyn:2019aio, Polyakov:2018rew}:
\begin{equation}
\label{eq:Mt}
M_n^{k_1 ...~k_n} (s', s) =
\int d^3r~r^n~Y_n^{k_1 ...~k_n}~
		T^{00}(\boldsymbol{r}, s', s) .
\end{equation}
The monopole contribution of this tensor quantity for a spin-3/2 particle is given by,
\begin{equation}
\label{eq:Mt0}
M_0 (s', s) =
\int d^3r~Y_0~\varepsilon_0 (r)~ \delta_{s' s}
= m \int d^3r~
[\varepsilon_0 (t)]_{FT}~
\delta_{s' s}
= m F_{1,0} (0)~
\delta_{s' s},
\end{equation}
where the normalization condition imposes the constraint $F_{1,0} (0) = 1$~\cite{Polyakov:2018zvc, Kim:2020lrs}.
	
The $0k$-components of the EMT describe the distribution of angular momentum within the system. The corresponding spin density is expressed as~\cite{Polyakov:2002yz, Polyakov:2018zvc, Kim:2020lrs},
\begin{align}\label{eq:EMTspace0i}
		\begin{aligned}
			J^i(\boldsymbol{r}, s', s) =
			\epsilon^{ijk} r^j 
			T^{0k}(\boldsymbol{r}, s', s) &= 
			2 \hat{S}^j_{s' s}
			{\Big[\Big( \mathcal{J}_1(t) 
				+ \frac{2}{3} t \frac{d \mathcal{J}_1(t)}{dt}
				\Big) Y_0 \delta^{ij}\Big]}_{FT} 
			- 2 \hat{S}^j_{s' s} {\Big[ t \frac{d \mathcal{J}_1(t)}{dt} Y^{ij}_2
				\Big]}_{FT}\\
			&+
			\frac{2}{m^2} \hat{O}^{jmn}_{s' s}
			{\Big[ t^2 \frac{d \mathcal{J}_3(t)}{dt} Y_4^{imnj} -  \Big( 2 t \mathcal{J}_3(t) 
				+ \frac{4}{7} t^2 \frac{d \mathcal{J}_3(t)}{dt}
				\Big)  \delta^{ij} Y_2^{mn}
				\Big]}_{FT}, 
		\end{aligned}
\end{align}
which includes distinct multipole contributions. 
The first and second terms in Eq.\eqref{eq:EMTspace0i} represent the monopole ($J^i_{\text{mono}}$) and quadrupole ($J^i_{\text{quad}}$) components of the spin density, respectively, analogous to the decomposition found in the nucleon case~\cite{Polyakov:2002yz, Polyakov:2018zvc}.
This quadrupole contribution, which is considerably smaller than the monopole term, has been discussed extensively in Refs.~\cite{Lorce:2017wkb, Schweitzer:2019kkd} and reflects the breaking of spherical symmetry down to axial symmetry induced by the polarization of the state. Importantly, upon integration over the entire spatial volume, this contribution vanishes and therefore does not affect the total angular momentum~\cite{Lorce:2017wkb, Schweitzer:2019kkd}. The averaged angular momentum density of the spin-3/2 baryon, based on its monopole angular momentum distribution, is expressed as~\cite{Kim:2020lrs}
\begin{equation}
\label{eq:J0}
		\rho_J(r)/S = \frac{1}{\text{Tr}[\hat{S}^2]}
		\sum_{s', s, i} \hat{S}^i_{s' s} J^i_{\text{mono}} (\boldsymbol{r}, s', s) = 
		-r \frac{d}{dr} [\mathcal{J}_1(t)]_{FT}, 
\end{equation}
where $S$ denotes the spin of the particle. The total angular momentum is determined by integrating over all space:
\begin{equation}
\label{eq:totalspin}
		\frac{1}{S} \int d^3 r \rho_J(r) = 2 \mathcal{J}_1(0) = \frac{2}{3} F_{4,0} (0)= 1. 
\end{equation}
which ensures proper normalization when both quark and gluon contributions are included. This yields the constraint $F_{4,0} (0)= \frac{3}{2}$.
	
The spatial $ij$-components of the energy-momentum tensor encode information about the internal mechanical properties of the hadron, specifically the pressure and shear force distributions. These components can be decomposed into multipole contributions associated with these densities as follows~\cite{Polyakov:2019lbq, Sun:2020wfo, Panteleeva:2020ejw, Kim:2020lrs}:
\begin{align}\label{eq:EMTspaceij}
\begin{aligned}
			T^{ij}(\boldsymbol{r}, s', s)
			&=
			p_0(r) \delta^{ij} \delta_{s' s} + s_0(r) Y_2^{ij} \delta_{s' s} 
			+ \Big(p_2(r) + \frac{1}{3} p_3(r) - \frac{1}{9} s_3(r) \Big) \hat{Q}^{ij}_{s' s}\\
			&+ \Big(s_2(r) - \frac{1}{2} p_3(r) + \frac{1}{6} s_3(r) \Big) 2\Big[\hat{Q}^{ik}_{s' s} Y_2^{kj} + \hat{Q}^{jk}_{s' s} Y_2^{ki} - \delta^{ij} \hat{Q}^{km}_{s' s} Y_2^{km}\Big]\\
			&+ \hat{Q}^{km}_{s' s} Y_2^{km} \Big[\Big(\frac{2}{3}p_3(r) + \frac{1}{9}s_3(r)\Big) \delta^{ij} + \Big(\frac{1}{2}p_3(r) + \frac{5}{6}s_3(r)\Big)Y_2^{ij}\Big], 
\end{aligned}
\end{align}
The monopole terms $p_0(r)$ and $s_0(r)$ describe the pressure and shear force distributions in the spherically symmetric systems. The functions $p_2(r)$ and $p_3(r)$ represent the quadrupole components of the pressure distribution, while $s_2(r)$ and $s_3(r)$ characterize the quadrupole components of the shear force distribution. These mechanical densities are defined as follows for $n=0,2,3$~\cite{Kim:2020lrs}:
\begin{eqnarray}
\label{eq:pressure}
		p_n(r) &=& \frac{1}{6m} \partial^2 \tilde{D}_n(r) =
		\frac{1}{6m} \frac{1}{r^2}
		\frac{d}{dr} r^2 \frac{d}{dr} \tilde{D}_n(r),
\\
\label{eq:shear}
		s_n(r) &=& -\frac{1}{4m} r 
		\frac{d}{dr} \frac{1}{r} \frac{d}{dr} \tilde{D}_n(r),
\end{eqnarray}
with	
\begin{eqnarray}
\label{eq:D0r}
		\tilde{D}_0(r) &=& [D_0(t)]_{FT},
		\\
		\label{eq:D2r}
		\tilde{D}_2(r) &=& [D_2(t)]_{FT} + \frac{1}{m^2}\Big(\frac{d}{dr}\frac{d}{dr} - \frac{2}{r} \frac{d}{dr}\Big) [D_3(t)]_{FT},
		\\
		\label{eq:D3r}
		\tilde{D}_3(r) &=& -\frac{2}{m^2}\Big(\frac{d}{dr}\frac{d}{dr} - \frac{3}{r} \frac{d}{dr}\Big) [D_3(t)]_{FT}.
\end{eqnarray} 
The conservation of the EMT for the static stress tensor,
$\nabla^i T_{ij}(r) = 0$, leads to the following differential equation, which relates the multipole components $p_n(r)$ and $s_n(r)$:
\begin{equation}
\label{eq:diff}
\frac{d}{dr}\Big(p_n(r) + \frac{2}{3} s_n(r)\Big) + \frac{2}{r} s_n(r) = 0.
\end{equation}
For spin-3/2 systems, the spherical components of the strong force acting on an infinitesimal radial area element 
$d \boldsymbol{S} = d S_r \hat{r} + d S_{\theta} \hat{\theta} + d S_{\phi} \hat{\phi}$ are given by~\cite{Panteleeva:2020ejw}:
\begin{eqnarray}
\frac{d F_r}{d S_r} &=& \Big( p_0(r) + \frac{2}{3} s_0(r) \Big)+ \hat{Q}^{rr} \Big(
p_2(r) + \frac{2}{3} s_2(r) +
p_3(r) + \frac{2}{3} s_3(r) \Big), \label{eq:dFr} \\
\frac{d F_{\theta}}{d S_r} &=&  \hat{Q}^{\theta r} \Big(
p_2(r) + \frac{2}{3} s_2(r) \Big), \qquad
\frac{d F_{\phi}}{d S_r} =  \hat{Q}^{\phi r} \Big(
p_2(r) + \frac{2}{3} s_2(r) \Big).
\label{eq:dFt}
\end{eqnarray}
While the force on the radial area element in spherically symmetric spin-0 and spin-1/2 hadrons is purely radial (normal)~\cite{Polyakov:2018zvc}, as described by the leading term in Eq.~\eqref{eq:dFr}, unpolarized spin-3/2 hadrons exhibit additional tangential components due to quadrupole spin polarization. 
Eq.~\eqref{eq:dFt} indicates that the tangential forces arise from $p_2(r)$ and $s_2(r)$. We define the longitudinal force associated with the $n$-th multipole contribution of the pressure and shear force as,
\begin{equation}
\label{eq:LF}
F^{||}_n(r) = p_n (r) + \frac{2}{3} s_n (r), \quad n=0, 2, 3. 
\end{equation}
The generalized D-terms $\mathcal{D}_{0,2,3}$ can be defined in terms of the spatial form factors $\tilde{D}_n(r)$, and, based on Eq.~\eqref{eq:diff}, also in terms of the multipole pressure and shear force components as~\cite{Panteleeva:2020ejw}:
\begin{equation}
\label{eq:Dtermsr}
		\mathcal{D}_n \equiv
		\int d^3 r \tilde{D}_n(r) = m \int d^3 r ~r^2~ p_n(r)
		= -\frac{4}{15} m \int d^3 r ~r^2 ~s_n(r), 
\end{equation}
or alternatively, in terms of the momentum-space form factors $D_{n}(t)$~\cite{Kim:2020lrs},
\begin{eqnarray}
\label{eq:Dtermst}
		\mathcal{D}_0 &=& D_0(0), \nonumber \\
		\mathcal{D}_2 &=& D_2(0) + \frac{2}{m^2}\int^{0}_{-\infty} dt D_3(t), \nonumber \\
		\mathcal{D}_3 &=& - \frac{5}{m^2} \int^{0}_{-\infty} dt D_3(t).
\end{eqnarray}
The generalized D-terms are dimensionless observables that capture the elastic behavior and mechanical stability of hadrons.

In this section, we discussed the GFFs and the associated mechanical properties of a spin-3/2 baryon in a model-independent framework. In the following section, we turn to the QCD sum rule method to explicitly calculate the GFFs of the $\Omega^{-}$ baryon.

\section{QCD sum rules}\label{sec:formalism}
	
We determine the gravitational form factors associated with the $\Omega^- \to \Omega^-$ transition by evaluating the following three-point correlation function within the framework of QCD sum rules,
\begin{equation}
\label{eq:corrf}
\Pi_{\alpha\mu\nu\beta}(p,q) = i^2 \int d^4 x e^{-ip.x} \int d^4 y e^{ip'.y}
		\langle 0 |\mathcal{T}[J_{\alpha}^{\Omega}(y)T_{\mu \nu}(0)\bar{J}_{\beta}^{\Omega}(x)]| 0 \rangle.
\end{equation}
In this expression, $p$ ($p'$) denotes the initial (final) four-momentum of the $\Omega^-$ baryon, $q = p' - p$ indicates the momentum transfer, $\mathcal{T}$ is the time ordering operator, and $J_{\alpha}^{\Omega}(y)$ represents the interpolating current of the $\Omega^-$ baryon, defined as \cite{Azizi:2016hbr},
\begin{equation}
\label{eq:interpolating}
J_{\alpha}^{\Omega}(y) = \epsilon_{abc} 
		\Big(s^{aT} (y) C\gamma_{\alpha} s^{b} (y)\Big) s^{c} (y) ,
\end{equation}
where $s(y)$ is identified as the strange quark field, $C$ denotes the charge conjugation operator, and $a$, $b$, and $c$ represent color indices. The correlation function in Eq.~\eqref{eq:corrf} involves the energy-momentum tensor 
$T_{\mu \nu}(0)$, defined in Eq.\eqref{eq:totalEMT} with its quark and gluon parts given in Eqs.~\eqref{eq:EMTcurrentquark} and~\eqref{eq:EMTcurrentgluon}. To handle the quark derivative terms, the EMT current is evaluated at a general point $z$ and the limit 
$z \to 0$ is taken.
In the Fock-Schwinger gauge, defined by 
$z^\mu A^a_\mu(z)=0$, the gluon field can be related to the field strength tensor as,
\begin{equation}
\label{eq:gluonfield}
A^{a}_{\mu}(z) = \int_{0}^{1}d\alpha \,\alpha \,
		z_\xi \,G_{\xi\mu}^{a}(\alpha z)
		= \frac{1}{2}z_{\xi} \, G_{\xi\mu}^{a}(0)
		+ \frac{1}{3}z_\eta \, z_\xi \, {D}_\eta \,
		G_{\xi\mu}^{a}(0)+\cdots.
\end{equation}
This relation implies that in the limit $z \to 0$, the gluon field vanishes, reducing the covariant derivatives in Eq.~\eqref{eq:derivdetail} to ordinary partial derivatives.

The QCD sum rule technique is fundamentally based on the evaluation of the correlation function in two distinct representations: the physical (phenomenological) side and the QCD (theoretical) side. The gravitational form factors are extracted by equating the coefficients of corresponding Lorentz structures from both representations. In the following, we detail the construction of the correlation function on each side.

\subsection{Physical side}\label{subsec:physical side}
	
The physical side of the correlation function is formulated in terms of hadronic degrees of freedom. To derive this representation, two complete sets of intermediate states—corresponding to the initial and final $\Omega^-$ baryons—are inserted into Eq.~\eqref{eq:corrf}~\cite{Khodjamirian:2020btr, Colangelo:2000dp}:
\begin{equation} \label{eq:CompeletSet}
		1=\vert 0\rangle\langle0\vert +\sum_{h}\int\frac{d^4 p_h}{(2\pi)^4}(2\pi) \delta(p^2_h-m^2)|h(p_h)\rangle  \langle h(p_h)|+\mbox{higher Fock states},
\end{equation}
where $m=m_{\Omega^-}$ and the summation runs over all hadronic states carrying the same quantum numbers as the $\Omega^-$ baryon. This completeness relation is applied separately for both the initial and final baryons with momenta $p_h$ and $p'_h$, respectively. Integrating over the four-dimensional space-time coordinates $x$ and $y$, and employing the identity~\cite{Dehghan:2025ncw},
\begin{equation} 
\label{eq:dxdp}
		\int d^4 x\int\frac{d^4 p_h}{(2\pi)^4}(2\pi) \delta(p^2_h-m^2)
		e^{i(p_h - p).x}
		= \frac{i}{m^2 - p^2},
\end{equation}
for both the initial and final baryon contributions isolates the ground-state pole terms. After performing the necessary calculations, the hadronic representation of the correlation function is obtained as,
\begin{equation}
\label{eq:physicalcorr}
\Pi_{\alpha\mu\nu\beta}^\text{Had}(p,q) = 
\sum_{s{'}}\sum_{s} \frac{\langle0|J_{\alpha}^{\Omega} (0) |{\Omega(p',s')}\rangle\langle {\Omega(p',s')}
|T_{\mu \nu}(0)|
			{\Omega(p,s)}\rangle\langle {\Omega(p,s)}
			|\bar{J}_{\beta}^{\Omega} (0) | 0 \rangle}
		{(m^2-p'^2)(m^2-p^2)} 
		+\cdots,
\end{equation}
where the dots denote the contributions from excited and continuum states. The first and third matrix elements in the numerator of Eq.~\eqref{eq:physicalcorr} are parameterized in terms of the residue of the $\Omega^-$ baryon, denoted by $\lambda_{\Omega}$, as follows:
\begin{equation}
\label{eq:residue}	\langle0|J_{\alpha}^{\Omega} (0)|{\Omega(p',s')}\rangle = \lambda_{\Omega} u_{\alpha}(p',s').
\end{equation}
The second matrix element in Eq.~\eqref{eq:physicalcorr} corresponds to the energy-momentum tensor current evaluated between $\Omega^-$ states and is parameterized in terms of the gravitational form factors, as presented in Eq.~\eqref{eq:matrix element}. Since we consider both quark and gluon contributions to the EMT, this results in seven independent and conserved gravitational form factors. By substituting the matrix elements from Eqs.~\eqref{eq:residue} and~\eqref{eq:matrix element} into Eq.~\eqref{eq:physicalcorr}, and employing the following completeness relation for the Rarita-Schwinger spinor of the $\Omega^-$ baryon,
\begin{equation}
\label{eq:sspin}
		\sum_{s'}{u_\alpha}(p',s') \bar{u}_{\alpha'}(p',s') = 
		- ({\pslash'} + m)
		\Big[g_{\alpha\alpha'} - \frac{\gamma_{\alpha}\gamma_{\alpha'}}{3}
		-\frac{2 p'_{\alpha} p'_{\alpha'}}{3 m^2}
		+\frac{p'_{\alpha} \gamma_{\alpha'} - p'_{\alpha'} \gamma_{\alpha}}{3 m}
		\Big],
\end{equation}
we obtain the hadronic representation of the correlation function in terms of the $\Omega^-$ baryon’s GFFs:
\begin{align}
\label{eq:physicalcorrfun}
\begin{aligned}			\Pi_{\alpha\mu\nu\beta}^\text{Had}&(p,q) = 
			\frac{-\lambda_{\Omega}^2}{(m^2-p'^2)(m^2-p^2)}
			({\pslash'} + m)
			\Big[g_{\alpha\alpha'} - \frac{\gamma_{\alpha}\gamma_{\alpha'}}{3}
			-\frac{2 p'_{\alpha} p'_{\alpha'}}{3 m^2}
			+\frac{p'_{\alpha} \gamma_{\alpha'} - p'_{\alpha'} \gamma_{\alpha}}{3 m}
			\Big]
			\\
			&\times\Big\{
			\frac{P_{\mu}P_{\nu}}{m}
			\Big(g^{\alpha' \beta'} F_{1,0}(t) 
			- \frac{\Delta^{\alpha'}\Delta^{\beta'}}{2 m^2} 
			F_{1,1}(t)\Big)
			+
			\frac{(\Delta_{\mu}\Delta_{\nu} - g_{\mu\nu} \Delta^2)}{4 m}
			\Big(g^{\alpha' \beta'} F_{2,0}(t) 
			- \frac{\Delta^{\alpha'}\Delta^{\beta'}}{2 m^2} 
			F_{2,1}(t)\Big)
			\\
			&+
			\frac{i}{2}
			\frac{(P_{\mu} \sigma_{\nu\rho} + P_{\nu} \sigma_{\mu\rho})
				\Delta^{\rho}}{m}
			\Big(g^{\alpha' \beta'} F_{4,0}(t) 
			- \frac{\Delta^{\alpha'}\Delta^{\beta'}}{2 m^2} 
			F_{4,1}(t)\Big)
			- \frac{1}{m}\Big(
			g^{\alpha'}_{\mu}\Delta_{\nu}\Delta^{\beta'} 
			+ g^{\alpha'}_{\nu}\Delta_{\mu}\Delta^{\beta'}
			\\
			&+
			g^{\beta'}_{\mu}\Delta_{\nu}\Delta^{\alpha'}
			+g^{\beta'}_{\nu}\Delta_{\mu}\Delta^{\alpha'}
			-2 g_{\mu\nu} \Delta^{\alpha'}\Delta^{\beta'} 
			- \Delta^2 g^{\alpha'}_{\mu} g^{\beta'}_{\nu}
			- \Delta^2 g^{\alpha'}_{\nu} g^{\beta'}_{\mu}
			\Big)F_{5,0}(t)\Big\}\\
			&\times
			({\pslash} + m)
			\Big[g_{\beta'\beta} - \frac{\gamma_{\beta'}\gamma_{\beta}}{3}
			-\frac{2 p_{\beta'} p_{\beta}}{3 m^2}
			+\frac{p_{\beta'} \gamma_{\beta} - p_{\beta} \gamma_{\beta'}}{3 m}
			\Big]+\cdots.  
\end{aligned}
\end{align}
The correlation function is constructed to describe the physical properties of the $\Omega^-$ baryon, a spin-3/2 state, through its gravitational form factors.
However, the Lorentz structures appearing in this expression are not all linearly independent, which complicates the extraction of physical observables, and the correlation function also includes contributions from spin-1/2 states.
The matrix element of the current $J_{\alpha}$ between the vacuum and a spin-1/2 baryon state is given by,
\begin{equation}
\label{spin12}
		\langle0\mid J_{\alpha}(0)\mid p,s=1/2\rangle=(A  p_{\alpha}+B\gamma_{\alpha})u(p,s=1/2).
\end{equation}
The spin-1/2 contributions thus appear through the structures $p_\alpha$ and $\gamma_\alpha$. Contracting both sides with $\gamma^\alpha$ and imposing the constraint $\gamma^\alpha J_\alpha = 0$ allows $A$ to be expressed in terms of $B$, revealing their linear dependence. To suppress these contributions and isolate independent Lorentz structures relevant to the spin-3/2 state in the correlation function, we adopt a specific Dirac matrix ordering: $\gamma_{\alpha} \qslash \pslash \gamma_{\mu} \gamma_{\nu} \gamma_{\beta}$. With this choice, and considering Eq.~\eqref{spin12} for the initial and final states with momenta $p$ and $p'$, we discard terms starting with $\gamma_\alpha$, ending with $\gamma_\beta$, and those proportional to $p_\beta$ and $p'_{\alpha}$ (with the substitution $p_{\alpha} \to -q_\alpha$), thereby eliminating spin-1/2 contaminations.
In the next step, we perform a double Borel transformation with respect to $p^2$ and $p'^2$, using a common Borel parameter $M^2$ and setting $M_i^2 = M_f^2 = 2 M^2$ for the identical initial and final $\Omega^-$ states. After continuum subtraction to suppress
contributions from higher states and continuum, the final hadronic representation of the correlation function is obtained,
\begin{align}\label{eq:HadStr}
\begin{aligned}
\Pi_{\alpha\mu\nu\beta}^\text{Had}&(Q^2) = \lambda_{\Omega}^2 e^{-\frac{m^2}{M^2}} \Big[ 
\Pi_{1}^\text{Had} (Q^2) 
p_{\mu} p_{\nu} g_{\alpha\beta} \mathbb{1}
			+ \Pi_{2}^\text{Had} (Q^2) 
			p_{\mu} p_{\nu} g_{\alpha\beta} \pslash
			+\Pi_{3}^\text{Had} (Q^2) 
			q_{\alpha} p_{\mu} p_{\nu} q_{\beta} \mathbb{1}
			+ \Pi_{4}^\text{Had} (Q^2)
			q_{\alpha} p_{\mu} p_{\nu} q_{\beta} \pslash \\
			&+ \Pi_{5}^\text{Had} (Q^2) 
			q_{\alpha} p_{\mu} p_{\nu} q_{\beta} \qslash
			+ \Pi_{6}^\text{Had} (Q^2) 
			p_{\mu} p_{\nu} g_{\alpha\beta} \qslash \pslash 
			+ \Pi_{7}^\text{Had} (Q^2) 
			q_{\alpha} q_{\mu} q_{\beta} \qslash \pslash {\gamma_{\nu}} 
			+\cdots\Big],
\end{aligned}
\end{align}
where $Q^2 = -t$, $\mathbb{1}$ denotes the unit matrix and the functions $\Pi_{i}^\text{Had} (Q^2)$ depend on the gravitational form factors. For clarity and brevity, we present only a subset of the full set of structures. We proceed by evaluating the correlation function on the QCD side.

\subsection{QCD side}\label{subsec:QCD side}

The QCD side representation of the correlation function is derived by substituting the interpolating current of the $\Omega^-$ baryon, given in Eq.~\eqref{eq:interpolating}, and the explicit form of the EMT current from Eq.~\eqref{eq:totalEMT}, which includes the quark and gluon contributions specified in Eqs.~\eqref{eq:EMTcurrentquark} and~\eqref{eq:EMTcurrentgluon}, into the definition of the correlation function in Eq.~\eqref{eq:corrf}. Subsequently, Wick’s theorem is applied to the resulting expression, and all possible contractions are evaluated, yielding the QCD form of the correlation function as follows:
\begin{align}
\begin{aligned}
\label{eq:contractions}	
\Pi_{\alpha\mu\nu\beta}^\text{QCD}&(p,q) = i^2 
	\epsilon^{abc} \epsilon^{a'b'c'} 
			\int  d^4 x \int d^4 y
			e^{-ip.x}  e^{ip'.y} \\
			&\times \Bigg\{ i \Bigg[
			2 S^{cb'}_{s}(y-x) \gamma_{\beta}
			S'^{ba'}_{s}(y-x) \gamma_{\alpha}
			S^{am}_{s}(y-z) 
			\Big(
			\gamma_{\nu}
			\overleftrightarrow{D}_\mu (z) 
			- g_{\mu\nu} \big( \overleftrightarrow{\dslash} + i m_s \big )
			\Big)
			S^{mc'}_{s}(z-x)\\
			& + 2 
			S^{cm}_{s}(y-z) 
			\Big(
			\gamma_{\nu}
			\overleftrightarrow{D}_\mu (z) 
			- g_{\mu\nu} \big( \overleftrightarrow{\dslash} + i m_s \big )
			\Big)
			S^{mb'}_{s}(z-x) \gamma_{\beta}
			S'^{ba'}_{s}(y-x) \gamma_{\alpha}
			S^{ac'}_{s}(y-x) 
			\\
			&+ 2 S^{cc'}_{s}(y-x) Tr \Big[ \gamma_{\beta} S'^{bb'}_{s}(y-x) \gamma_{\alpha} S^{am}_{s}(y-z) 
			\Big(
			\gamma_{\nu}
			\overleftrightarrow{D}_\mu (z) 
			- g_{\mu\nu} \big( \overleftrightarrow{\dslash} + i m_s \big )
			\Big)
			S^{ma'}_{s}(z-x) \Big] \\
			&+ S^{cm}_{s}(y-z) 
			\Big(
			\gamma_{\nu}
			\overleftrightarrow{D}_\mu (z) 
			- g_{\mu\nu} \big( \overleftrightarrow{\dslash} + i m_s \big )
			\Big)
			S^{mc'}_{s}(z-x)
			Tr \Big[ \gamma_{\beta} S'^{bb'}_{s}(y-x) \gamma_{\alpha}S^{aa'}_{s}(y-x) \Big]
			\\
			&+2 S^{ca'}_{s}(y-x) \gamma_{\beta}
			S'^{mb'}_{s}(z-x)
			\Big(
			\gamma_{\nu}
			\overleftrightarrow{D}_\mu (z) 
			- g_{\mu\nu} \big( \overleftrightarrow{\dslash} - i m_s \big )
			\Big)
			S'^{am}_{s}(y-z) \gamma_{\alpha} S^{bc'}_{s}(y-x) 
			+ \mu \leftrightarrow \nu 
			\Bigg]\\
			&+ 
			\langle G^2 \rangle g_{\mu\nu} 
			\Bigg[ 
			2 S^{ca'}_{s}(y-x) \gamma_{\beta} S'^{ab'}_{s}(y-x)
			\gamma_{\alpha}S^{bc'}_{s}(y-x)
			+ S^{cc'}_{s}(y-x) Tr \Big[ \gamma_{\beta} S'^{aa'}_{s}(y-x)
			\gamma_{\alpha} S^{bb'}_{s}(y-x) \Big]
			\Bigg]
			\Bigg\},
\end{aligned}
\end{align}
where $m_s$ represents the mass of the strange quark, $S' = C S^T C$, and $S^{ij}_s(x)$ denotes the propagator of the strange quark, with the explicit form given by: 
\begin{align}
\label{eq:lightquarkProp}
		S^{ij}_s(x) &= i \delta_{ij} \dfrac{\xslash}{2\pi^2 x^4} - 
		\delta_{ij} \dfrac{m_s}{4 \pi^2 x^2} - \delta_{ij} \dfrac{\langle\bar{s}s\rangle}{12} 
		+ i \delta_{ij} \dfrac{\xslash m_s \langle\bar{s}s\rangle}{48}
		- \delta_{ij} \dfrac{x^2}{192} m_0^2 \langle\bar{s}s\rangle 
		+ i \delta_{ij} \dfrac{x^2 \xslash m_s}{1152} m_0^2 \langle\bar{s}s\rangle 
		\nonumber\\
		&
		- i \dfrac{g_s G_{ij}^{\lambda\delta}}{32 \pi^2 x^2} 
		[\xslash \sigma_{\lambda\delta} + \sigma_{\lambda\delta} \xslash]
		+ \cdots,
\end{align}
where ${\langle \bar{s} s\rangle}$ corresponds to the strange quark condensate, and $m_0^2 = \langle\bar{s}g_s G^{\mu\nu}\sigma_{\mu\nu}s\rangle / \langle\bar{s}s\rangle$ characterizes the quark-gluon condensate. 
In Eq.~\eqref{eq:contractions}, the two main square brackets arise from the quark and gluon components of the EMT current, respectively. Inserting the explicit forms of the strange quark propagators and applying the covariant derivatives in Eq.~\eqref{eq:contractions}, and subsequently taking the limit $z\rightarrow 0$, yields:
\begin{align}
\label{eq:qcd}
\Pi_{\alpha\mu\nu\beta}^{\text{QCD}}(p,q) = \int d^4 x \int d^4 y e^{-ip.x}  e^{ip'.y} 
		\Big\{ 
		\Big[
		\Gamma_{\alpha\mu\nu\beta}^{(P)}
		+ \Gamma_{\alpha\mu\nu\beta}^{(3D)}
		+ \Gamma_{\alpha\mu\nu\beta}^{(4D,q)}
		+ \Gamma_{\alpha\mu\nu\beta}^{(5D)} 
		+ \mu \leftrightarrow \nu
		\Big]
		+ \Gamma_{\alpha\mu\nu\beta}^{(4D,g)}
		\Big\}.
\end{align}
The QCD side correlation function presented above consists of a perturbative contribution, denoted by $\Gamma^{(P)}$, along with non-perturbative contributions $\Gamma^{(3D)}$, $\Gamma^{(4D)}$, and $\Gamma^{(5D)}$, which correspond to the three, four, and five dimensions, respectively. The explicit expressions for these terms are provided in Appendix~\ref{Calcu}. The terms enclosed in square brackets originate from the quark part of the EMT, while the term outside the brackets, $\Gamma^{(4D,g)}$, emerges from the gluonic part of the EMT current. The four-dimensional non-perturbative contributions from the quark and gluon sectors, given in Eqs.~\eqref{eq:4Dquark} and~\eqref{eq:4Dgluon}, contain products of two gluon field strength tensors, $G^{A}_{\alpha\beta}$; these products are responsible for the gluon condensate and are thoroughly discussed in Appendix B of Ref.~\cite{Dehghan:2023ytx}.
In the next step, we transform the QCD side correlation function from coordinate space to momentum space and apply Feynman parameterization to facilitate the evaluation of the resulting integrals. After performing the necessary calculations and simplifications, the correlation function is expressed in terms of double dispersion integrals:
\begin{equation}
\label{eq:spectral density}
		\Pi_{i}^{\text{QCD}}(Q^2) = \int_{3 m_s^2}^{s_0} ds \int_{3 m_s^2}^{s_0} ds'
		\frac{\rho_i(s,s',Q^2)}{(s-p^2)(s'-p'^2)},
\end{equation}
where $s_0$ denotes the continuum threshold. Since the initial and final particles are identical 
$\Omega^-$ baryons, the lower and upper limits of the double integrals coincide. The spectral densities $\rho_i(s,s',Q^2)$ are determined by the imaginary parts of $\Pi_{i}^{\text{QCD}}(Q^2)$ through the relation $\rho_i(s,s',Q^2) = Im[\Pi_{i}^{\text{QCD}}(Q^2)]/\pi$. Details of these QCD side calculations can be found in our previous analysis of the $\Delta$ baryon in Ref.~\cite{Dehghan:2023ytx}; to avoid redundancy, they are not presented here. Finally, by imposing the same Dirac matrix ordering as on the physical side, removing spin-1/2 contaminations, and applying the double Borel transformation and continuum subtraction to the QCD representation of the correlation function, we obtain,
	\begin{align}
		\label{eq:QCDStr}
		\Pi_{\alpha\mu\nu\beta}^\text{QCD}(Q^2) &= \int_{3 m_s^2}^{s_0} ds \int_{3 m_s^2}^{s_0} ds' 
		e^{-s/{2M^2}} e^{-s'/{2M^2}}
		\Big[ 
		\Pi_{1}^\text{QCD} (Q^2, s, s')
		p_{\mu} p_{\nu} g_{\alpha\beta} \mathbb{1}
		+ \Pi_{2}^\text{QCD} (Q^2, s, s')  
		p_{\mu} p_{\nu} g_{\alpha\beta} \pslash \nonumber\\
		&+ \Pi_{3}^\text{QCD} (Q^2, s, s') 
		q_{\alpha} p_{\mu} p_{\nu} q_{\beta} \mathbb{1}
		+ \Pi_{4}^\text{QCD} (Q^2, s, s') 
		q_{\alpha} p_{\mu} p_{\nu} q_{\beta} \pslash
		+ \Pi_{5}^\text{QCD} (Q^2, s, s') 
		q_{\alpha} p_{\mu} p_{\nu} q_{\beta} \qslash \nonumber\\
		&+ \Pi_{6}^\text{QCD} (Q^2, s, s')
		p_{\mu} p_{\nu} g_{\alpha\beta} \qslash \pslash 
		+ \Pi_{7}^\text{QCD} (Q^2, s, s') 
		q_{\alpha} q_{\mu} q_{\beta} \qslash \pslash {\gamma_{\nu}} + \cdots\Big].
\end{align}
Equating the structures obtained from the hadronic and QCD sides allows us to extract the explicit analytical expressions of the $\Omega^-$ baryon’s gravitational form factors. The next section presents these form factors and their numerical analysis.

\section{Numerical analysis and GFFs}\label{sec:numerical}
	
This section provides a detailed numerical analysis of the gravitational form factors of the $\Omega^-$ baryon, determined from the QCD sum rules established in the preceding sections. Throughout the analysis, we adopt a renormalization scale of $\mu = 1$ GeV, which is commonly used in QCD sum rule calculations for light hadrons and lies within the non-perturbative regime, consistent with the standard values of input condensates. The input parameters used in the analysis, including the strange quark mass, quark and gluon condensate values, and other relevant quantities, are listed in Table~\ref{table:inputPar}. The sum rule calculations also involve two auxiliary parameters—the Borel parameter $M^2$ and the continuum threshold $s_0$—and, in principle, physical observables should remain independent of these parameters. Since complete independence is not achievable, we identify stability regions where the gravitational form factors show minimal sensitivity to these parameters. 
To ensure the reliability of the QCD sum rule analysis, the Borel parameter $M^2$ and the continuum threshold $s_0$ are chosen such that the pole contribution (PC) dominates over the higher states and continuum. We require,
\begin{equation}
\label{eq:PC}
\text{PC}(Q^2) = \frac{\Pi_{i}^{\text{QCD}}(Q^2, M^2, s_0)}{\Pi_{i}^{\text{QCD}}(Q^2, M^2, \infty)} \geqslant 0.5,
\end{equation}
where $i$ denote the specific Lorentz structures. This criterion guarantees that at least half of the total QCD contribution originates from the ground-state pole contribution. This restriction fixes the upper limit of the Borel parameter $M^2$. Its lower bound is determined by the convergence of the operator product expansion (OPE). In the present analysis, non-perturbative contributions are included up to dimension 5. To ensure OPE convergence, we impose the condition,
\begin{equation}
\label{eq:R}
\text{R} (M^2, Q^2) = \frac{\Pi_{i}^{\text{QCD, Dim5}}(Q^2, M^2, s_0)}{\Pi_{i}^{\text{QCD}}(Q^2, M^2, \infty)} \leqslant 0.08.
\end{equation}
This requirement guarantees that the highest-dimensional term retained in the OPE contributes no more than $8\%$ of the total QCD result. 
Applying these constraints, the following working windows for $M^2$ and $s_0$ are derived from our analyses:
\begin{align}
\label{eq:working region}
		5.5 ~\text{GeV}^2 &\leqslant M^2 \leqslant 6.5 ~\text{GeV}^2,\\
		\nonumber
		3.3 ~\text{GeV}^2 &\leqslant s_0 \leqslant 3.5 ~\text{GeV}^2.
\end{align}

\begin{table}[!htb]
\centering	
		\begin{tabular}{c*{12}{c}r}	
			\hline
			\hline
			$m_s$ 
			& \qquad $96^{+8}_{-4}$ MeV~\cite{Aliev:2016jnp}
			& \qquad $m_{\Omega^-}$ 
			& \qquad
			$1657 \pm 172~\mathrm{MeV}$~\cite{Aliev:2016jnp} 
			& \qquad $\lambda_{\Omega^-}$ 
			& \qquad $0.049 \pm 0.015~\mathrm{GeV}^3$~\cite{Aliev:2016jnp}
			\\
			$\langle \bar{s}s \rangle$ 
			& \qquad $0.8 (-0.24\pm 0.01)^3$ $\mathrm{GeV}^3$~\cite{Aliev:2016jnp} 
			& \qquad $m_{0}^2$ 
			& \qquad $(0.8 \pm 0.1)$ $\mathrm{GeV}^2$~\cite{Belyaev:1982sa}
			& \qquad $\alpha_s$ 
			& \qquad $0.118 \pm 0.005$~\cite{DELPHI:1993ukk}
			\\
			$\langle \frac{\alpha_s}{\pi} G^2 \rangle$ 
			& \qquad $(0.012\pm0.004)$ $~\mathrm{GeV}^4$~\cite{Belyaev:1982cd}
			& \qquad $g_s^2$ 
			& \qquad $4 \pi \alpha_s$
			\\
			\hline
\end{tabular}
\caption{Key input parameters and their values used in the analysis.}
\label{table:inputPar}
\end{table}

We calculate the numerical values of the gravitational form factors of the $\Omega^-$ baryon in the momentum transfer squared range $0 ~\text{GeV}^2 \leqslant -t \leqslant 10 ~\text{GeV}^2$ within the QCD sum rules framework. Fig.~\ref{fig:GFFs} illustrates the behavior of the form factors in the low momentum transfer region for better clarity. A commonly adopted approach in the analysis of gravitational form factors involves employing the $p$-pole parametrization, a fitting technique that offers a reliable method for extrapolating the behavior of the form factors beyond the calculated momentum transfer region. The following $p$-pole function is utilized to fit the obtained gravitational form factors:
\begin{figure}[!htb]
		\centering
		\includegraphics[scale=0.6]{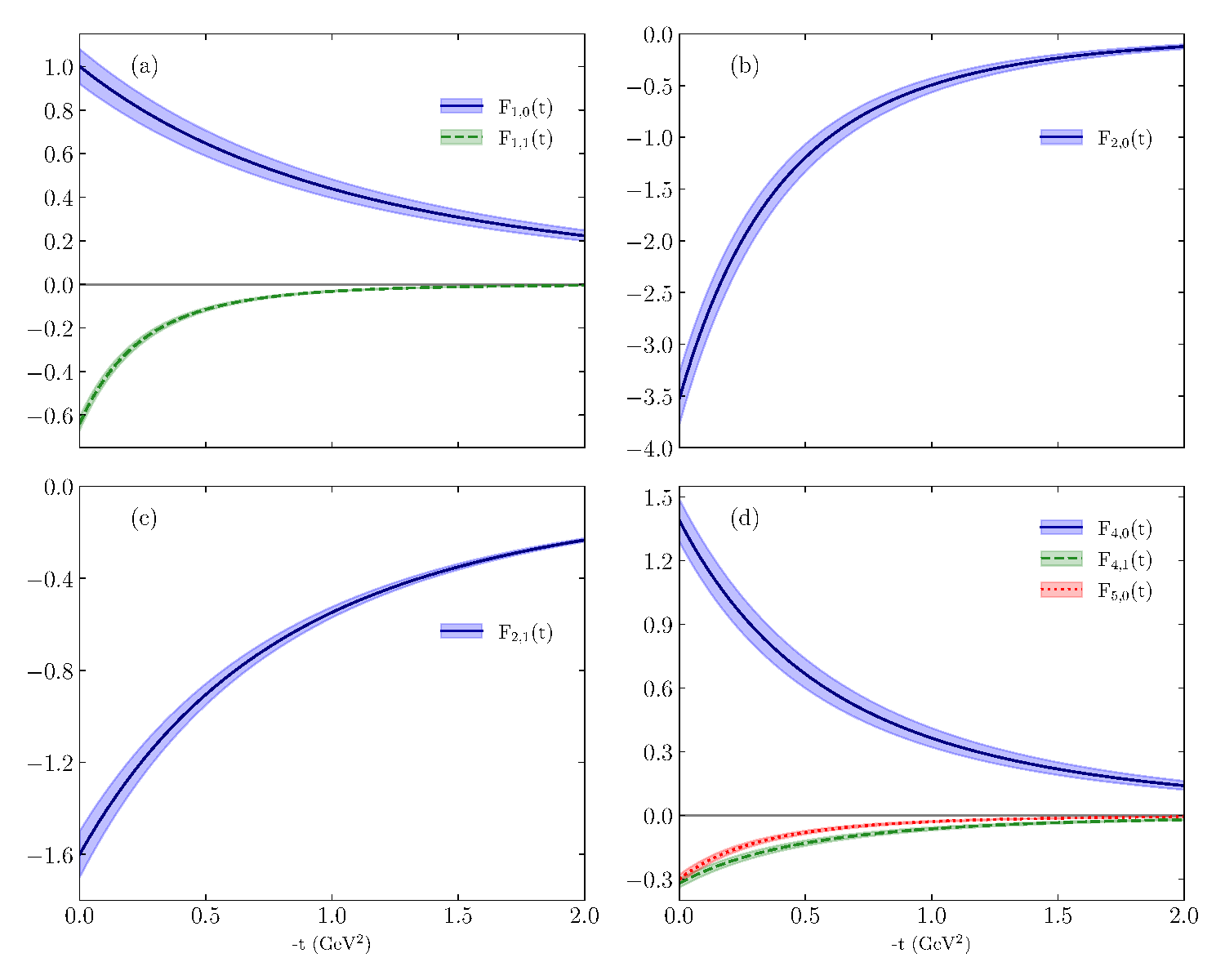}
\caption{\small 
The seven conserved gravitational form factors of the $\Omega^-$ baryon as functions of momentum transfer squared $-t$, with shaded bands indicating uncertainties. A $p$-pole fit is applied.}
\label{fig:GFFs}
\end{figure}
\begin{equation}\label{eq:FitFun}
		\mathcal{G}(t) = \frac{{\mathcal{G}}(0)}{\Big(1- g_{p}\,t\Big)^p},
\end{equation}
where ${\cal G} (0)$ and $p$ are dimensionless fit parameters, while $g_{p}$ carries the dimension of inverse energy squared (GeV$^{-2}$). The $p$-pole fit parameters corresponding to the gravitational form factors shown in Fig.~\ref{fig:GFFs} are presented in Table~\ref{table:fitparameters}.
The uncertainties in the extracted form factors arise from variations within the chosen ranges of the helping parameters $M^2$ and $s_0$, inaccuracies in the input parameters, and the inherent systematic errors of the QCD sum rules approach.
Importantly, the values of the gravitational form factors at zero momentum transfer, ${\cal G} (0)$, reflect essential mechanical properties of the $\Omega^-$ baryon, such as the mass, spin and D-terms. The precise determination of these quantities thus plays a crucial role in advancing our understanding of hadron structure from first principles.

\begin{table}[!htb]
\centering	
		\begin{tabular}{c*{3}{c} | c*{12}{c}}	
			\hline
			\hline
			GFF & \qquad ${\cal G}(0)$  & \quad $g_{p}$~(GeV$^{-2}$) & 
			\quad 
			\hspace{0.6 cm} 
			$p$ 
			\hspace{1. cm}
			& 
			\quad
			GFF & \qquad ${\cal G}(0)$  & \quad $g_{p}$~(GeV$^{-2}$) & \quad $p$
			\\
			\hline
			$F_{1,0}(t)$ & \qquad $1.00(\pm8)$  &\quad $0.25(\mp1)$ &\quad $3.70(\pm5)$ 
			& 
			\quad
			$F_{1,1}(t)$ & \qquad $-0.64(\mp 3)$  &\quad$0.80(\mp 2)$ &\quad $5.10(\pm 3)$  \\
			$F_{2,0}(t)$ & \qquad $-3.53(\mp 25)$ &\quad$0.51(\mp 2)$ &\quad$4.77(\mp 1)$
			& 
			\quad
			$F_{2,1}(t)$ & \qquad $-1.60(\mp 10)$  &\quad$0.29(\pm 1)$ &\quad $4.21(\mp 5)$  \\
			$F_{4,0}(t)$ & \qquad $1.39(\pm 10)$  &\quad$0.49(\mp 3)$ &\quad $3.36(\pm 4)$
			& 
			\quad
			$F_{4,1}(t)$ & \qquad $-0.32(\mp 2)$  &\quad$0.52(\mp 3)$ &\quad $3.90(\pm 5)$  \\
			$F_{5,0}(t)$ & \qquad $-0.30(\mp 2)$  &\quad$0.72(\mp 2)$ &\quad $4.31(\mp 1)$
			\\
			\hline
\end{tabular}
\caption{\small 
The $p$-pole fit parameters ${\cal G}(0)$, $g_{p}$, and $p$ for the GFFs shown in Fig.~\ref{fig:GFFs}. }
\label{table:fitparameters}
\end{table}

The impact of the $p$-pole parameterization of the form factor on the mechanical distributions should be emphasized.
The mechanical distributions are extracted from the energy–momentum tensor in coordinate space, which involves a three-dimensional Fourier transform of the form factors. As shown explicitly in Eq.~\eqref{eq:fourier}, this requires integrating the form factors over the full space-like momentum transfer region ($ t \leqslant 0$) to obtain the complete spatial distributions.
In our approach, the QCD sum rule framework provides reliable results for the gravitational form factors in the domain $0 ~\text{GeV}^2 \leqslant -t \leqslant 10 ~\text{GeV}^2$, while the extrapolation to asymptotically large momentum transfer is implemented through a $p$-pole parametrization. As a result, the spatial mechanical distributions and the corresponding radii are influenced by the assumed large-$t$ behavior and should be regarded as consequences of the $p$-pole ansatz constrained by our results in the fitting region. In particular, the asymptotic behavior ($t \to -\infty$) of the form factors may have a significant impact on the shape of the mechanical distribution at small $r$. By contrast, quantities determined directly from the form factors at $t=0$ are fixed by the QCD sum rules, whereas the detailed $r$-dependence of the mechanical densities requires a $p$-pole parametrization.

Although lattice QCD studies have not yet provided results for the $\Omega^-$ GFFs, our findings can serve as theoretical benchmarks for forthcoming lattice simulations and experimental measurements. In the subsequent section, we use the gravitational form factors extracted here to explore the mechanical properties of the $\Omega^-$ baryon. These properties, encoded in the momentum transfer dependence of the form factors, offer a unique perspective on the internal dynamics and stability of hadrons—revealing information that goes beyond what can be accessed through electromagnetic or axial form factor studies alone.

\section{Mechanical properties}\label{sec:mech}

\begin{figure}[!htb]
		\centering
		\includegraphics[scale=0.6]{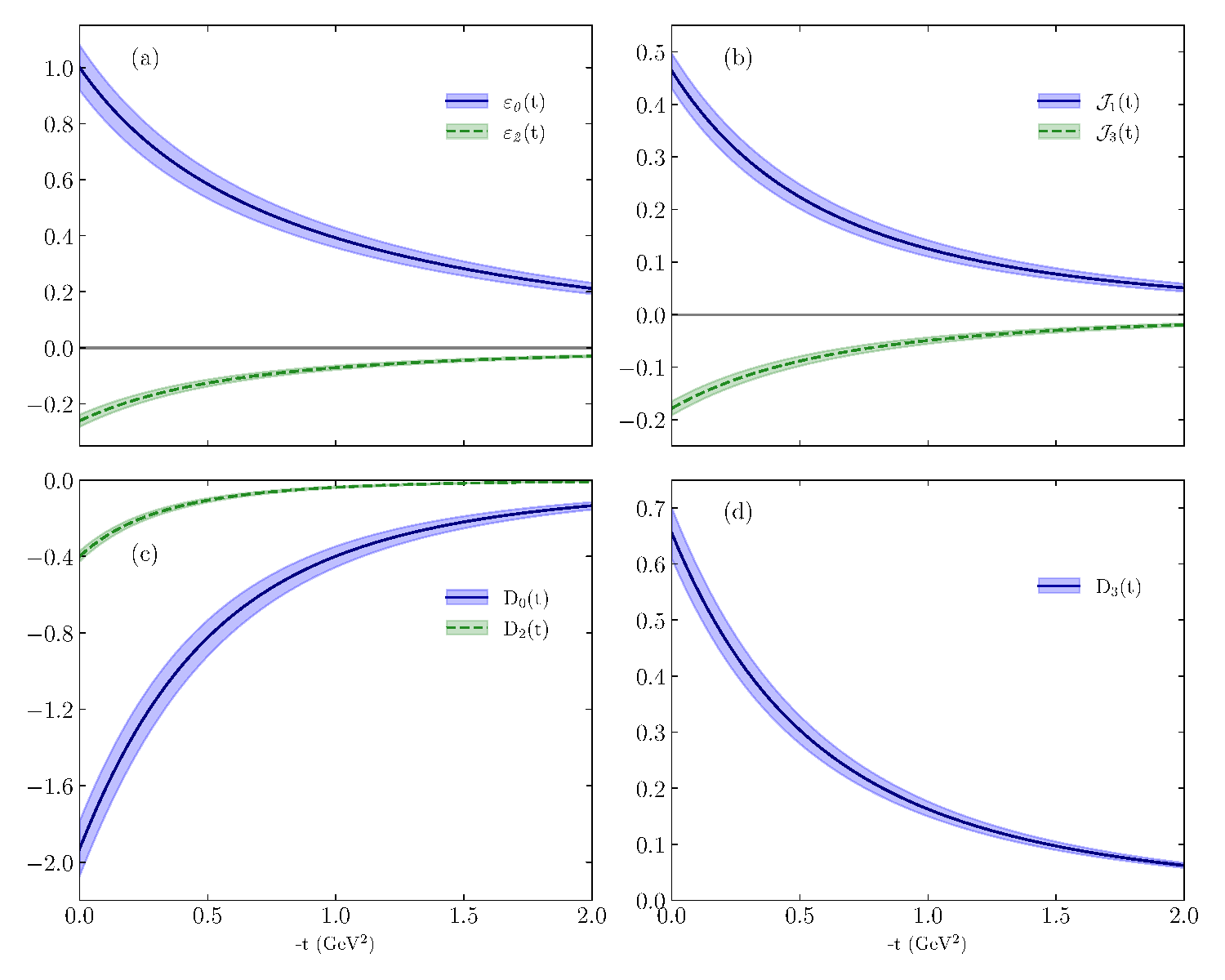}
		\caption{\small 
The GMFFs of the $\Omega^-$ baryon as functions of momentum transfer squared $-t$, shown in the range $0 ~\text{GeV}^2 \leqslant -t \leqslant 2 ~\text{GeV}^2$ for clarity.}
\label{fig:GMFFs}
\end{figure}

In this section, we investigate the mechanical characteristics of the $\Omega^-$ baryon, including its internal energy density, angular momentum, pressure, and shear force distributions, as determined by its gravitational form factors. The spin-3/2 nature of the $\Omega^-$ gives rise to higher-order multipole contributions in these distributions, reflecting the richer internal structure and complex spatial patterns of the stress-energy tensor components.
In the Breit frame, where a three-dimensional spatial interpretation of the energy-momentum tensor becomes possible, the gravitational multipole form factors (GMFFs) emerge in the matrix elements of the EMT in momentum space, detailed in Eqs.~\eqref{eq:matrix element00},~\eqref{eq:matrix element0i}, and~\eqref{eq:matrix elementij}. These expressions introduce monopole, dipole, and higher-order GMFFs, which encapsulate the multipole structure of the mechanical distributions. The GMFFs, defined in Eq.~\eqref{eq:GMFFs}, are constructed as specific combinations of the GFFs. 
Fig.~\ref{fig:GMFFs} shows the GMFFs of the $\Omega^-$ baryon extracted from our calculations as functions of the squared momentum transfer $-t$, including their corresponding uncertainties, within the range
$0 ~\text{GeV}^2 \leqslant -t \leqslant 2 ~\text{GeV}^2$.
Panel (a) displays the energy-monopole $\varepsilon_0(t)$ and energy-quadrupole $\varepsilon_2(t)$ form factors. Panel (b) shows the angular momentum-dipole $\mathcal{J}_1(t)$ and angular momentum-octupole $\mathcal{J}_3(t)$ form factors. The remaining multipole form factors, $D_0(t)$, $D_2(t)$, and $D_3(t)$, are depicted in panels (c) and (d).
It is observed that, among the multipole contributions to the energy, angular momentum, and $D(t)$ form factors, the lowest-order multipole component in each case dominates.
The GFFs of the $\Omega^-$ baryon and its mechanical properties have been investigated within a covariant quark-diquark framework in Refs.~\cite{Fu:2023ijy, Wang:2023bjp}. Our results for $\varepsilon_0(t)$ and $\mathcal{J}_1(t)$ are consistent with those studies, while our determined distributions for $\varepsilon_2(t)$ and $D_0(t)$ exhibit different signs compared to those reported in Refs.~\cite{Fu:2023ijy, Wang:2023bjp}.
The analysis of our $\varepsilon_2(t)$ through Eq.~\eqref{eq:energyr2} leads to a positive spatial quadrupole energy density, $\varepsilon_2(r)$ (see Fig.~\ref{fig:physdensity}), which is consistent with fundamental physical principles. Since $\varepsilon_2(r)$ corresponds to a mass density distribution, its positivity ensures a positive mass contribution associated with the quadrupole deformation of the $\Omega^-$ baryon. From Eq.~\eqref{eq:Dtermst}, we obtain $\mathcal{D}_{0} = D(0) = -1.93(14)$, where the negative sign of the D-term is in accordance with the mechanical stability condition of the baryon.

The values of the GMFFs at zero momentum transfer represent key intrinsic mechanical characteristics of the $\Omega^-$ baryon, including its mass, spin, and D-terms. These values are summarized in Table~\ref{table:GMFFsZerot}.
Our result, $\varepsilon_0(0) = F_{1,0} (0) = 1.00(8)$, is in excellent agreement with the mass normalization condition detailed in Section~\ref{sec:general}~\cite{Polyakov:2018zvc, Kim:2020lrs}.
The dipole contribution to the angular momentum at $t=0$ is found to be $\mathcal{J}_1(0) = F_{4,0}(0)/3 = 0.46(3)$, where the extracted value $F_{4,0}(0) = 1.39(10)$ is consistent with the expected spin 3/2 of the $\Omega^-$ baryon. The generalized D-terms, which encode information about the internal mechanical forces and stability of the baryon, are defined in Eq.~\eqref{eq:Dtermst}; a comprehensive discussion follows later in this section.

\begin{table}[!htb]
		\centering	
		\begin{tabular}{c*{12}{c}r}	
			\hline\hline
$\varepsilon_0(0)$ &\quad 
			$
			\varepsilon_2(0) 
			$
			&\quad
			$\mathcal{J}_1(0)$ &\quad 
			$
			\mathcal{J}_3(0)
			$
			&\quad
			$D_0(0)$ &\quad 
			$D_2(0)$ &\quad $D_3(0)$  \\
			\hline
$ 1.00(8) $ &\quad $-0.26(2)$ &\quad
$0.46(3)$ &\quad $-0.18(1)$ &\quad
$-1.93(14)$ &\quad $-0.40(3)$ &\quad $0.66(5)$ 
			\\
			\hline
		\end{tabular}
		\caption{The values of the $\Omega^-$ GMFFs at zero momentum transfer. }
		\label{table:GMFFsZerot}
\end{table}

\begin{figure}[!htb]
\centering
\includegraphics[scale=0.6]{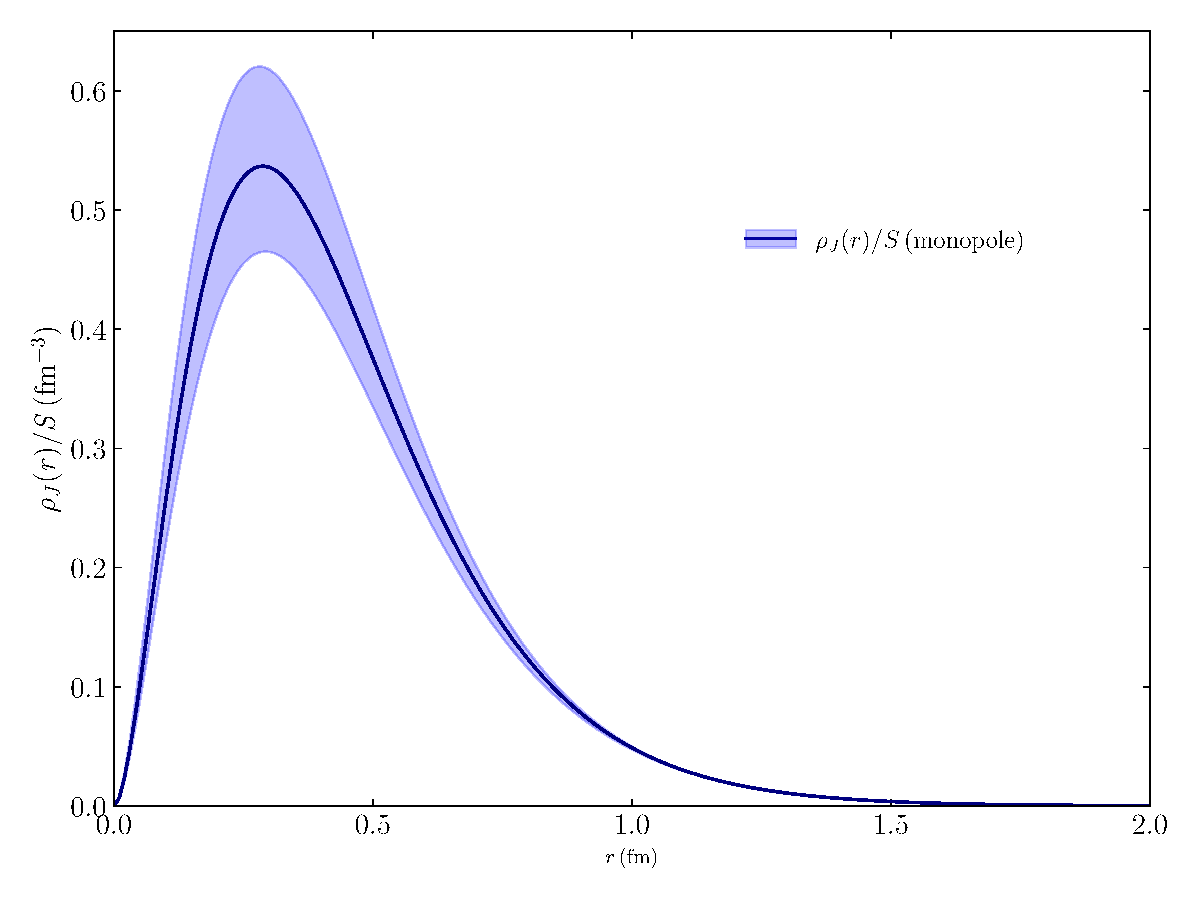}
\caption{\small 
Monopole contribution to the spatial distribution of the angular momentum density in the $\Omega^-$ baryon, normalized by its spin $S=3/2$. }
\label{fig:Jdensity}
\end{figure}

\begin{figure}[!htb]
\includegraphics[scale=0.6]{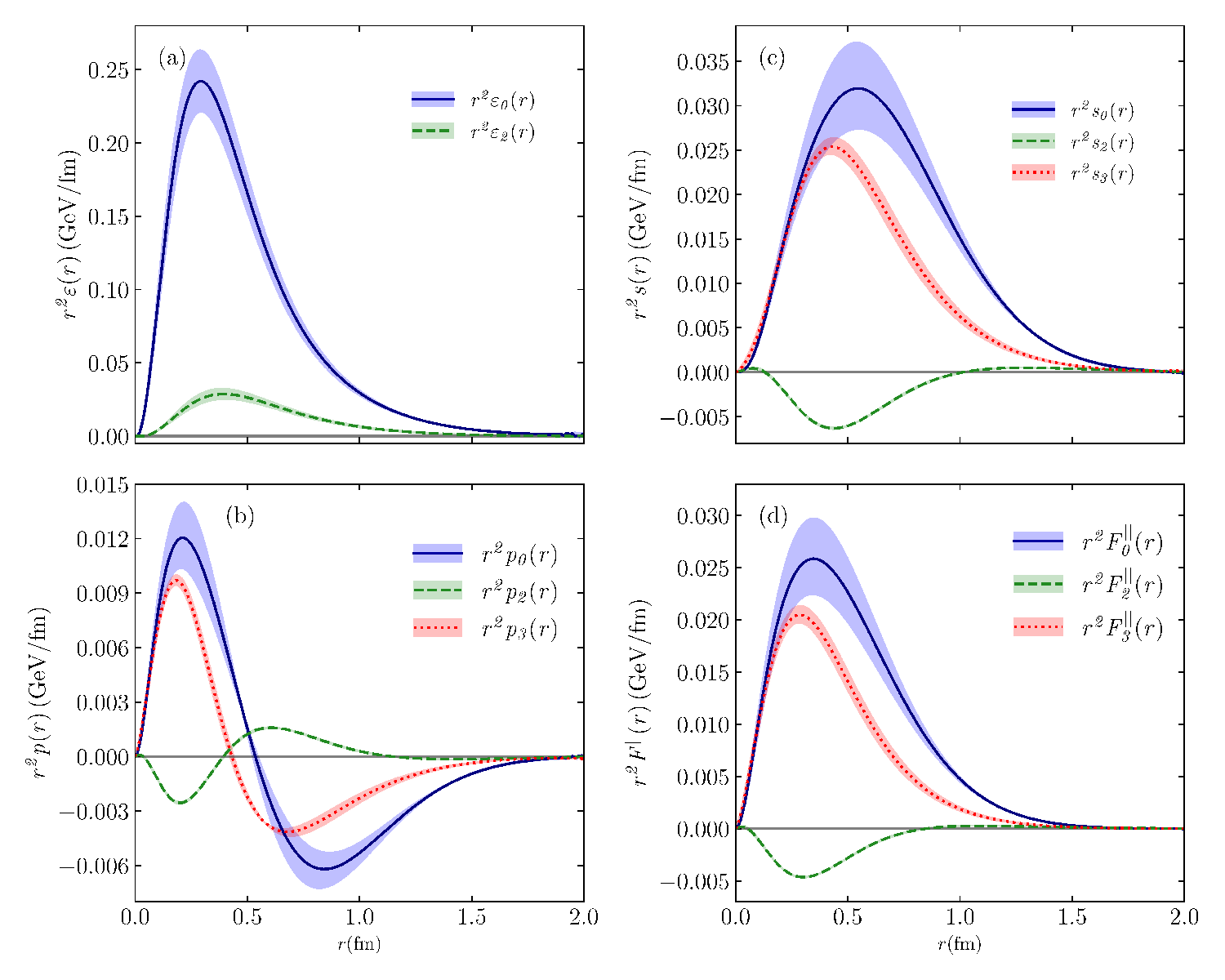}
\caption{\small
The mutipole contributions to the spatial distributions of the energy density, pressure, shear force, and longitudinal force in the $\Omega^-$ baryon. }
\label{fig:physdensity}
\end{figure}

Having established the GMFFs in momentum space, we apply three-dimensional Fourier transforms to obtain spatial distributions of energy density, angular momentum density, pressure, and shear forces within the $\Omega^-$ baryon. These distributions enable detailed exploration of its internal mechanical structure.
The angular momentum density of the $\Omega^-$ baryon includes multipole contributions, as shown in Eq.~\eqref{eq:EMTspace0i}, where the quadrupole term is significantly smaller than the monopole component and integrates to zero over the entire spatial volume, leaving the monopole contribution as the dominant contributor to the total angular momentum~\cite{Lorce:2017wkb, Schweitzer:2019kkd}.
Fig.~\ref{fig:Jdensity} displays the normalized monopole contribution of the angular momentum density, $\rho_J(r)/S$, defined in Eq.~\eqref{eq:J0}, as a function of the radial distance $r$. 
This distribution provides direct insight into the internal spin structure of the baryon, with the angular momentum density reaching its maximum at $r \approx 0.3\,\mathrm{fm}$. Our calculations yield a total angular momentum of $\int d^3 r \rho_J(r) = 1.39 (10)$, which agrees with the expected spin value of $S=3/2$.

We determine the monopole and quadrupole contributions to the spatial distributions of the energy density, pressure, shear force, and longitudinal force inside the $\Omega^-$ baryon by employing Eqs.~\eqref{eq:energyr0},~\eqref{eq:energyr2},~\eqref{eq:pressure},~\eqref{eq:shear}, and~\eqref{eq:LF}, respectively. These distributions are shown in Fig.~\ref{fig:physdensity}, where for the energy density, $n=0$ and $n=2$ correspond to the monopole and quadrupole terms, respectively, while for the pressure, shear force, and longitudinal force, $n=0$ denotes the monopole and $n=2, 3$ the quadrupole components. 
In all of these spatial distributions, the monopole component emerges as the dominant contribution.
For the pressure, shear force, and longitudinal force distributions, the quadrupole component with $n=2$ generally displays the smallest contribution.
Furthermore, the $n=2$ distributions are approximately inverse to the $n=3$ distributions, such that the peaks observed in the $n=3$ components roughly coincide with the valleys in the $n=2$ components. Notably, in the pressure distributions, the first node of the $n=2$ component nearly coincides with that of the $n=3$ component, indicating a characteristic correlation between these quadrupole contributions.

Fig.~\ref{fig:physdensity}~(a) displays the spatial distributions of the monopole and quadrupole components of the energy density inside the $\Omega^-$ baryon.
Interestingly, the monopole energy density exhibits a peak at $r \approx 0.3\,\mathrm{fm}$, similar to the monopole angular momentum density shown in Fig.~\ref{fig:Jdensity}, indicating that both the mass and the spin are concentrated within the same central region.
The quadrupole component of the energy density has the same positive sign as the monopole component and peaks at $r \approx 0.4\,\mathrm{fm}$. However, its magnitude at all radial distances is considerably smaller than that of the monopole energy density. The typical sizes of the monopole and quadrupole energy densities can be estimated by integrating them over all space,
\begin{eqnarray}
\label{eq:E0}
	\int d^3r~\varepsilon_0(r) &=& 1657(135)~\text{MeV},\\
\label{eq:E2}
	\int d^3r~\varepsilon_2(r) &=& 225(24)~\text{MeV}.
\end{eqnarray}
A comparison of Eqs.~\eqref{eq:Mt0} and~\eqref{eq:E0} with the value of $m_{\Omega^-}$ in Table~\ref{table:inputPar} shows that the integral of $\varepsilon_0(r)$ accurately reproduces the mass of the $\Omega^-$ baryon in our calculations.
Moreover, the integral of $\varepsilon_2(r)$ amounts to approximately $14\%$ of the monopole contribution, highlighting the relatively small but non-negligible role of the quadrupole component in the baryon’s energy distribution.

The gravitational form factors $D_{0} (t)$, $D_{2} (t)$, and $D_{3} (t)$ play a central role in revealing the internal dynamics and geometric properties of the $\Omega^-$ baryon. These form factors determine the multipole structures of the pressure and shear force distributions and define the generalized D-terms $\mathcal{D}_{0,2,3}$, which capture key information about the internal forces. Assessing whether these forces lead to a physically consistent configuration requires evaluating stability conditions, which describe how the internal pressures and forces must balance to prevent the baryon from expanding or collapsing. A fundamental global stability criterion, derived from the conservation of the EMT, is the von Laue condition,
\begin{equation}
\label{eq:von Laue}
\int_{0}^{\infty} r^2 dr ~ p(r)= 0,
\end{equation}
which requires the pressure distribution to change sign at least once~\cite{Polyakov:2018zvc} and guarantees an overall balance of internal pressure. 
However, this global constraint alone cannot confirm full mechanical stability, since unstable states may also satisfy it~\cite{Perevalova:2016dln}.
To provide a more stringent test, we examine two local stability conditions on the internal force distributions. The first condition requires the longitudinal force to satisfy,
\begin{equation}
\label{eq:stabilityF}
F_{||}(r)>0,
\end{equation}
which, in spin-1/2 systems, corresponds to an outward-directed force counteracting collapse. In spin-3/2 systems, however, the presence of quadrupole spin polarization introduces additional tangential force components into the normal force, so this condition should be interpreted more cautiously.
The second local condition requires the shear force distribution to remain positive,
\begin{equation}
\label{eq:stabilitys}
s(r)>0,
\end{equation}
reflecting hydrostatic equilibrium within the baryon. 
While positivity of $s(r)$ and $p(r) + 2/3 \,s(r)$ has been suggested for spin-1/2 systems such as the nucleon~\cite{Polyakov:2018zvc}, no formal proof exists for their validity in higher-spin systems. Indeed, it remains an open question whether these local stability criteria hold universally for spin-1/2 hadrons and higher-spin systems. Consequently, we regard these relations as phenomenological indicators of local stability rather than rigorous theorems, and we assess their applicability to the 
$\Omega^-$ baryon. 
According to the first local stability condition, the D-term must be negative, a fundamental quantity whose sign and magnitude are still debated~\cite{Polyakov:1999gs, Perevalova:2016dln, Burkert:2023wzr}.
The generalized D-terms $\mathcal{D}_{0,2,3}$ of the $\Omega^-$ baryon, defined via the spatial form factors $\tilde{D}_n(r)$ in Eq.~\eqref{eq:Dtermsr}, are determined as,
\begin{equation}
\label{eq:DtermsOmega}
\mathcal{D}_0 = -1.93(14), \qquad
\mathcal{D}_2 = 0.001(3), \qquad
\mathcal{D}_3 = -1.00(7)
\end{equation}
Furthermore, applying Eq.~\eqref{eq:Dtermst} confirms these results. It is noteworthy that the quadrupole component with $n=2$ is nearly zero, while the monopole component and the $n=3$ quadrupole component satisfy the stability conditions by exhibiting negative D-terms. Interestingly, this behavior mirrors what is found for the 
$\Delta$ baryon, which shows a similar pattern in its D-terms~\cite{Panteleeva:2020ejw, Kim:2020lrs, Dehghan:2025ncw}. 
For spin-3/2 particles, the vanishing of 
$\mathcal{D}_2$ was first identified in the chiral soliton model within the large-$N_c$ framework~\cite{Panteleeva:2020ejw}. In this approach, the quadrupole spatial distributions of pressure and shear force with $n=2$ vanish, i.e., 
$p_2(r) = s_2(r) = 0$, which, through Eq.~\eqref{eq:Dtermsr}, leads to $\mathcal{D}_2 = 0$. 
Remarkably, our QCD sum rule analysis for the $\Omega^-$ baryon yields a similar outcome $\mathcal{D}_2 \simeq 0$, while producing nonzero quadrupole components of the pressure and shear force densities, $p_2(r)$ and $s_2(r)$, as illustrated in Figs.~\ref{fig:physdensity}~(b) and~(c).

The behavior of the individual multipole components further elucidates the internal structure. Fig.~\ref{fig:physdensity}~(b) presents the spatial distribution of the monopole ($n=0$) and quadrupole ($n=2,3$) components of the internal pressure in the $\Omega^-$ baryon. Our results demonstrate that these pressure distributions satisfy the von Laue stability condition given in Eq.~\eqref{eq:von Laue}. This condition ensures the stability of the baryon through a balance between the outward-directed positive pressure and the inward-directed negative pressure.
For the monopole and $n=3$ quadrupole components—consistent with observations in other hadrons such as the nucleon~\cite{Polyakov:2018zvc, Dehghan:2025ncw}—the pressure is positive in the inner region and negative in the outer region. However, the $n=2$ quadrupole component exhibits an inverse pattern, with negative pressure in the inner region and positive pressure in the outer region. The nodes of the monopole and quadrupole pressure components are located approximately at $r \approx 0.55\,\mathrm{fm}$ and $r \approx 0.4\,\mathrm{fm}$, respectively. Fig.~\ref{fig:physdensity}~(c) shows the spatial distribution of the monopole and quadrupole components of the shear force. We observe that the monopole and $n=3$ quadrupole components satisfy the local stability condition $s(r) > 0$, indicating stable shear profiles that maintain the baryon’s integrity. In contrast, the $n=2$ component violates this condition by developing a region of negative shear force, signaling localized mechanical instability. Notably, our results for the monopole and the $n=2$ and $n=3$ quadrupole components of the pressure and shear force spatial distributions in the $\Omega^-$ baryon exhibit a pattern consistent with that found in a lattice QCD study of the $\Delta$ baryon based on the gluon energy-momentum tensor~\cite{Pefkou:2021fni}.
Fig.~\ref{fig:physdensity}~(d) presents the monopole and quadrupole contributions to the longitudinal force distributions. The local stability condition $F_{||}(r)>0$ is satisfied for the monopole and $n=3$ quadrupole components, whereas the $n=2$ quadrupole component fails this criterion. 
Although the $n=2$ quadrupole component violates the first and second local mechanical stability conditions, its contribution is subdominant compared to the monopole and $n=3$ components. Consequently, the $\Omega^-$ baryon remains mechanically stable.

The mass and mechanical radii of the $\Omega^-$ baryon are fundamental observables that offer valuable insight into its internal structure. The mass radius characterizes the spatial distribution of the energy density, while the mechanical radius, derived from the pressure and shear force distributions, reflects the extent of internal QCD forces acting within the baryon. Importantly, both the mass and mechanical radii receive contributions not only from the spherically symmetric monopole term but also from higher multipole components, which arise from the intrinsic spin-3/2 nature of the baryon. The mass and mechanical radii associated with different multipole components of the $\Omega^-$ baryon are defined as follows~\cite{Polyakov:2018zvc, Kim:2020lrs},
\begin{eqnarray}
\label{eq:massR}
	\langle r^2_n\rangle_{\text{mass}} &=&  
\frac{\int d^3r~ r^2 ~\varepsilon_n(r)}{\int d^3r ~\varepsilon_n(r)}, \qquad \text{with}~n=0, 2,\\
\nonumber\\
\label{eq:mechR}
	\langle r^2_n\rangle_{\text{mech}} &=& 
\frac{\int d^3r ~r^2 ~F_n^{||}(r)}{\int d^3r ~F_n^{||}(r)}, \qquad \text{with}~n=0, 2, 3.
\end{eqnarray}
The corresponding results obtained from our analysis are summarized in Table~\ref{table:radii}. The quadrupole mass radius, ${\langle r^2_2\rangle}^{1/2}_{\text{mass}} = 0.637(43)\,\mathrm{fm}$, exceeds the monopole mass radius, ${\langle r^2_0\rangle}^{1/2}_{\text{mass}} = 0.551(32)\,\mathrm{fm}$, indicating that the quadrupole energy density distribution is more spatially extended. Our result for the monopole mass radius is consistent with the value ${\langle r^2_0\rangle}^{1/2}_{\text{mass}} = 0.544\,\mathrm{fm}$ presented in Ref.~\cite{Fu:2023ijy}.
For the mechanical radii, the monopole and $n=3$ quadrupole components yield comparable values, ${\langle r^2_0\rangle}^{1/2}_{\text{mech}} = 0.571(40)\,\mathrm{fm}$ and ${\langle r^2_3\rangle}^{1/2}_{\text{mech}} = 0.503(28)\,\mathrm{fm}$, respectively. In contrast, the $n=2$ quadrupole mechanical radius is substantially smaller, with ${\langle r^2_2\rangle}^{1/2}_{\text{mech}}  =0.111(29)\,\mathrm{fm}$, indicating that the internal QCD forces associated with this component are highly localized. This striking suppression directly results from the suppressed spatial distributions of 
$p_2(r)$, $s_2(r)$, and the associated longitudinal force $F^{||}_2(r)$, as shown in Fig.~\ref{fig:physdensity}.
The monopole mass radius of the $\Omega^-$ is significantly smaller than our previously reported value for the $\Delta$ baryon, ${\langle r^2_0\rangle}^{1/2}_{\text{mass}} = 0.818(25)\,\mathrm{fm}$~\cite{Dehghan:2023ytx}. Additionally, Ref.~\cite{ Kim:2020lrs} reports ${\langle r^2_0\rangle}^{1/2}_{\text{mech}} = 0.922\,\mathrm{fm}$ and ${\langle r^2_3\rangle}^{1/2}_{\text{mech}} = 0.574\,\mathrm{fm}$ for the monopole and $n=3$ quadrupole mechanical radii of the $\Delta$, respectively. 
These comparisons indicate that both the mass and mechanical radii of the $\Omega^-$ are smaller than those of the $\Delta$ baryon. This observation is consistent with the trend reported in Table 6 of Ref.~\cite{Wang:2023bjp}, which shows that the mass radii of decuplet baryons systematically decrease as baryon mass increases.
Compared to the proton, both the monopole mass and mechanical radii of the $\Omega^-$ baryon are noticeably smaller, indicating a more compact spatial structure due to the presence of heavier strange quarks. 
In our previous study~\cite{Dehghan:2025ncw}, we found $\langle r^2_p\rangle^{1/2}_{\text{mass}} = 0.697(37)\,\mathrm{fm}$ and $\langle r^2_p\rangle^{1/2}_{\text{mech}} = 0.629(26)\,\mathrm{fm}$ for the proton mass and mechanical radii; see Figure 4 of that work, which shows results from other methods and approaches.
This pattern is consistent with trends observed in charge radii extracted from electromagnetic form factors, which suggest that the charge radii of decuplet baryons are generally smaller than those of the octet baryons~\cite{Boinepalli:2009sq}.

\begin{table}[!htb]
\centering	
\begin{tabular}{c*{12}{c}r}	
\hline\hline
\rule{0pt}{3.5ex}
$\sqrt{\langle r^2_0\rangle_{\text{mass}}}$ (fm)  &\quad 
$\sqrt{\langle r^2_2\rangle_{\text{mass}}}$ (fm) 
&\quad
$\sqrt{\langle r^2_0\rangle_{\text{mech}}}$ (fm) 
&\quad
$\sqrt{\langle r^2_2\rangle_{\text{mech}}}$ (fm) 
&\quad
$\sqrt{\langle r^2_3\rangle_{\text{mech}}}$ (fm) 
\rule[-2.5ex]{0pt}{0pt} 
\\
\hline
$ 0.551(32) $ &\quad $0.637(43)$ &\quad
$0.571(40)$
&\quad $0.111(29)$ &\quad
$0.503(28)$
\\
\hline
\end{tabular}
\caption{Multipole mass and mechanical radii of the $\Omega^-$ baryon. }
\label{table:radii}
\end{table}

\section{Conclusion}\label{sec:con}

In this work, we investigated the mechanical properties of the $\Omega^-$ baryon through the analysis of its gravitational form factors within the three-point QCD sum rules framework. 
By including both the quark and gluon components of the energy-momentum tensor, we extracted the complete set of seven conserved gravitational form factors. 
These form factors were used to construct the gravitational multipole form factors (GMFFs): $\varepsilon_0(t)$ and $\varepsilon_2(t)$, describing the energy distribution; $\mathcal{J}_1(t)$ and $\mathcal{J}_3(t)$, associated with angular momentum; and $D_0(t)$, $D_2(t)$, and $D_3(t)$, characterizing internal pressure and shear forces. From this analysis, we found that in each category, the lowest-order multipole component provides the dominant contribution.
The emergence of such multipole structures is a direct consequence of the spin-3/2 nature of the $\Omega^-$ baryon.
Employing Fourier transformations of the GMFFs in the Breit frame, we obtained the multipole structure of the spatial distributions associated with key mechanical quantities—including energy density, angular momentum, pressure, shear force, and longitudinal force—and determined corresponding multipole observables such as the mass and mechanical radii and D-terms.
While the monopole contributions were found to dominate most of these observables, higher-order multipole components introduce nontrivial deformations and break spherical symmetry in the internal structure, with the $n=2$ quadrupole contributions identified as subdominant yet non-negligible.

A particularly noteworthy result we obtained is the spatial localization of the $\Omega^-$ baryon’s spin and mass. The monopole component of the angular momentum density exhibits a pronounced peak at $r \approx 0.3\,\mathrm{fm}$, closely matching the peak of the energy density distribution. This overlap indicates that the spin and mass are concentrated within the same central region, reflecting a strong spatial correlation between internal mass and angular momentum generation. 
We examined the mechanical stability of the $\Omega^-$ baryon through a multipole analysis of its internal force distributions. By evaluating the pressure, shear force, and longitudinal force distributions for the monopole and quadrupole components, we verified that the global von Laue condition is satisfied, indicating overall pressure balance. At the local level, we found that both the monopole and $n=3$ quadrupole components fulfill the stability criteria—namely, positive shear and longitudinal forces—while the $n=2$ quadrupole component exhibits localized violations. This instability may also be reflected in its small generalized D-term, $\mathcal{D}_2 \simeq 0$, in contrast to the negative D-terms obtained for the monopole and $n=3$ quadrupole components, which are consistent with mechanical equilibrium. 
Additionally, we observed a notable correlation between the $n=2$ and $n=3$ quadrupole components in the pressure, shear force, and longitudinal force distributions:
their nodes tend to align, and the peaks of the $n=3$ components approximately coincide with the valleys of the $n=2$ components. This compensatory behavior may further suppress the destabilizing effects of the $n=2$ component.
Despite these local instabilities,
the dominance of the stable monopole and $n=3$ contributions ensures that the $\Omega^-$ baryon remains mechanically stable.

We analyzed the mass and mechanical radii of the $\Omega^-$ baryon across various multipole components—quantifying the spatial distributions of energy density and internal QCD forces, respectively—and found that these radii are consistently smaller than those of the proton and $\Delta$ baryons. 
This indicates a more compact internal structure, in line with the presence of heavier strange quarks. These results are in agreement with theoretical expectations and previous studies, which show that within the decuplet baryon family, mass radii decrease with increasing baryon mass.
Furthermore, we found that the $n=2$ quadrupole mechanical radius of the 
$\Omega^-$ baryon is significantly smaller than those of other components, reflecting highly localized internal QCD forces. Our analysis confirms that this localization arises directly from the suppressed spatial distributions of pressure, shear force, and the longitudinal force within this component.

The findings presented in this work advance our understanding of the gravitational structure and mechanical dynamics of spin-3/2 baryons. They provide valuable benchmarks for future theoretical investigations, lattice QCD simulations, and GPD-based experimental efforts aimed at probing the internal structure of the $\Omega^-$ and other higher-spin systems.

\begin{acknowledgments}
		
Z. Dehghan and K. Azizi are thankful to the Iran National Science Foundation (INSF) for the financial support provided for this research under project number 4025645.
		
\end{acknowledgments}
	
\appendix
	
\setcounter{equation}{0}
\renewcommand{\theequation}{\Alph{section}.\arabic{equation}}
	
\section{Parameters in the EMT multipole expansions: GMFFs and Operators}\label{sec:AppGMFFs}

In this appendix, we present the explicit definitions of the parameters that appear in the multipole expansion of the energy-momentum tensor (EMT) matrix elements for spin-$3/2$ baryons, as given in Eqs.~\eqref{eq:matrix element00},~\eqref{eq:matrix element0i}, and~\eqref{eq:matrix elementij}. These parameters include the gravitational multipole form factors (GMFFs), irreducible tensors, and multipole spin operators. 
	
In the Breit frame, the GMFFs of a spin-$3/2$ baryon can be expressed in terms of the invariant conserved gravitational form factors $F_{i,j}(t)$ with $i=1,2,4,5$, and are explicitly defined by~\cite{Kim:2020lrs}:
\begin{eqnarray}
		\label{eq:GMFFs}
		\varepsilon_0(t) &=& F_{1,0}(t)  
		+ \frac{t}{6 m^2} \bigg[-\frac{5}{2} F_{1,0}(t) - F_{1,1}(t) 
		-\frac{3}{2} F_{2,0}(t) + 4 F_{5,0}(t) + 3 F_{4,0}(t)
		\bigg] \nonumber\\
		&&+ \frac{t^2}{12 m^4} \bigg[\frac{1}{2} F_{1,0}(t) 
		+ F_{1,1}(t) + \frac{1}{2} F_{2,0}(t) + \frac{1}{2} F_{2,1}(t)
		- 4 F_{5,0}(t) - F_{4,0}(t) - F_{4,1}(t)
		\bigg]
		\nonumber\\
		&&+ \frac{t^3}{48 m^6} \bigg[-\frac{1}{2} F_{1,1}(t) 
		- \frac{1}{2} F_{2,1}(t) + F_{4,1}(t)
		\bigg],
		\\
		\varepsilon_2(t) &=& -\frac{1}{6} \bigg[ 
		F_{1,0}(t) + F_{1,1}(t) -4 F_{5,0}(t)
		\bigg] \nonumber\\
		&&+ \frac{t}{12 m^2} \bigg[\frac{1}{2} F_{1,0}(t) + F_{1,1}(t) 
		+\frac{1}{2} F_{2,0}(t) +\frac{1}{2} F_{2,1}(t) - 4 F_{5,0}(t) - F_{4,0}(t)
		- F_{4,1}(t)
		\bigg] \nonumber\\
		&&+ \frac{t^2}{48 m^4} \bigg[-\frac{1}{2} F_{1,1}(t) 
		- \frac{1}{2} F_{2,1}(t) + F_{4,1}(t)
		\bigg],
		\\
		\mathcal{J}_1(t) &=& \frac{1}{3} F_{4,0}(t) 
		- \frac{t}{15 m^2} \bigg[F_{4,0}(t) + F_{4,1}(t) 
		+ 5 F_{5,0}(t)
		\bigg] + \frac{t^2}{60 m^4} F_{4,1}(t) ,
		\\
		\mathcal{J}_3(t) &=& - \frac{1}{6} \bigg[F_{4,0}(t) + F_{4,1}(t)\bigg]
		+ \frac{t}{24 m^2} F_{4,1}(t) ,
		\\
		D_0(t) &=& F_{2,0}(t) - \frac{16}{3} F_{5,0}(t) 
		- \frac{t}{6 m^2} \bigg[F_{2,0}(t) + F_{2,1}(t) 
		- 4 F_{5,0}(t)
		\bigg] + \frac{t^2}{24 m^4} F_{2,1}(t),
		\\
		D_2(t) &=& \frac{4}{3} F_{5,0}(t),
		\\
		D_3(t) &=& \frac{1}{6} \bigg[-F_{2,0}(t) - F_{2,1}(t) 
		+ 4 F_{5,0}(t)
		\bigg] + \frac{t}{24 m^2} F_{2,1}(t).
\end{eqnarray} 
These form factors describe the distributions of energy, angular momentum, and mechanical properties (such as pressure and shear forces) inside the baryon.

The $n$-rank irreducible tensors in coordinate and momentum spaces are formulated by~\cite{Kim:2020lrs},
\begin{equation}
\label{eq:irr}
		Y^{i_1 \dots i_n}_n (\Omega_r) = 
		\frac{(-1)^n}{(2n-1)!!} r^{n+1} \partial^{i_1} \partial^{i_2} ...~ \partial^{i_n} \frac{1}{r}, \qquad
		Y^{i_1 \dots i_n}_n (\Omega_p) = 
		\frac{(-1)^n}{(2n-1)!!} p^{n+1} \partial^{i_1} \partial^{i_2} ...~ \partial^{i_n} \frac{1}{p}.
\end{equation}
The quadrupole and octupole spin-dependent operators, $\hat{Q}^{ij}$ and $\hat{O}^{ijk}$ respectively, are symmetric, traceless, and constructed from the spin operator $\hat{S}^i$ as~\cite{Kim:2020lrs},
\begin{eqnarray}
		\label{eq:spinOperator}
		\hat{Q}^{ij} &=& \frac{1}{2} 
		\Big(\hat{S}^i \hat{S}^j + \hat{S}^j \hat{S}^i - \frac{2}{3} S(S+1) \delta^{ij} \Big),\nonumber\\
		\hat{O}^{ijk} &=& \frac{1}{6} 
		\Big(\hat{S}^i \hat{S}^j \hat{S}^k + \hat{S}^j \hat{S}^i \hat{S}^k 
		+ \hat{S}^k \hat{S}^j \hat{S}^i
		+ \hat{S}^j \hat{S}^k \hat{S}^i
		+ \hat{S}^i \hat{S}^k \hat{S}^j
		+ \hat{S}^k \hat{S}^i \hat{S}^j \nonumber\\
		&-& \frac{6 S(S+1) - 2}{5} 
		(\delta^{ij} \hat{S}^k + \delta^{ik} \hat{S}^j + \delta^{kj} \hat{S}^i)
		\Big),
\end{eqnarray} 
where $i, j, k = 1, 2, 3$. The spin operator $\hat{S}^{\lambda}_{s' s}$ transforms as a rank-1 tensor (vector operator) under SU(2), corresponding to the spin-1 irreducible representation, and can be expressed in terms of Clebsch–Gordan coefficients in the spherical basis~\cite{Polyakov:2019lbq}:
\begin{equation}
\label{eq:Sop}
\hat{S}^{\lambda}_{s' s} = \sqrt{S(S+1)} \, \, C^{S s'}_{Ss, 1\lambda}.
\end{equation}
This structure reflects the coupling of a spin-$S$ particle state with a spin-1 vector operator:
\begin{equation}
\label{eq:Structure}
\underbrace{|S, s \rangle}_{\text{particle}}
\hspace{0.4cm}
\otimes \underbrace{|1, \lambda \rangle}_{\text{vector operator}}
 \longrightarrow \quad
\underbrace{|S, s' \rangle}_{\text{particle}},
\end{equation}
where $\lambda=0,  \pm 1$. For spin-3/2 particles, we set $S=3/2$, with 
$s, s' = \pm 1/2, \pm 3/2$. The Clebsch–Gordan coefficient appearing in Eq.~\eqref{eq:Sop} is given by~\cite{Varshalovich:1988ifq},
\begin{equation}
\label{eq:CG}
C^{j_3 m_3}_{j_1 m_1, j_2 m_2} = (-1)^{j_1 - j_2 + m_3}\sqrt{2 j_3 + 1}
\begin{pmatrix}
j_1 & j_2 & j_3 \\
m_1 & m_2 & -m_3
\end{pmatrix},
\end{equation}
where the quantity in parentheses denotes the Wigner $3 jm$ symbol~\cite{Varshalovich:1988ifq}.

\section{Components of the QCD side correlation function}\label{Calcu}

In this appendix, we present the explicit expressions for the perturbative and non-perturbative components of the QCD side correlation function, as introduced in Eq.~\eqref{eq:qcd}, 
\begin{widetext} 
\begin{align}\label{eq:perturbative}
\begin{aligned}
				\Gamma_{\alpha\mu\nu\beta}^{(P)} 
				&=
				\dfrac{3 i^3}{(2 \pi^2)^4} 
				\Bigg\{ 2 \Big( i \dfrac{\yslash - \xslash}{(y-x)^{4}} - \dfrac{m_s}{2(y-x)^{2}}  \Big)
				\gamma_{\beta} \Big( i \dfrac{\yslash - \xslash}{(y-x)^{4}} + \dfrac{m_s}{2(y-x)^{2}}  \Big) \gamma_{\alpha} E^P_{\mu\nu} (x,y)
				\\
				&+
				2 E^P_{\mu\nu} (x,y)
				\gamma_{\beta} \Big( i \dfrac{\yslash - \xslash}{(y-x)^{4}} + \dfrac{m_s}{2(y-x)^{2}}  \Big)
				\gamma_{\alpha} 
				\Big( i \dfrac{\yslash - \xslash}{(y-x)^{4}} - \dfrac{m_s}{2(y-x)^{2}}  \Big)
				\\	
				&+
				2 \Big( i \dfrac{\yslash - \xslash}{(y-x)^{4}} - \dfrac{m_s}{2(y-x)^{2}}  \Big)
				Tr \Big[
				\gamma_{\beta} \Big( i \dfrac{\yslash - \xslash}{(y-x)^{4}} + \dfrac{m_s}{2(y-x)^{2}}  \Big) \gamma_{\alpha} 
				E^P_{\mu\nu} (x,y)
				\Big]
				\\
				&+
				E^P_{\mu\nu} (x,y)
				Tr \Big[
				\gamma_{\beta} \Big( i \dfrac{\yslash - \xslash}{(y-x)^{4}} + \dfrac{m_s}{2(y-x)^{2}}  \Big)
				\gamma_{\alpha} 
				\Big( i \dfrac{\yslash - \xslash}{(y-x)^{4}} - \dfrac{m_s}{2(y-x)^{2}}  \Big)
				\Big]
				\\
				&+
				2 \Big( i \dfrac{\yslash - \xslash}{(y-x)^{4}} - \dfrac{m_s}{2(y-x)^{2}}  \Big)
				\gamma_{\beta} 
				F^P_{\mu\nu} (x,y)
				\gamma_{\alpha} 
				\Big( i \dfrac{\yslash - \xslash}{(y-x)^{4}} - \dfrac{m_s}{2(y-x)^{2}}  \Big)  \Bigg\},
			\end{aligned}
		\end{align}
	\end{widetext}

\begin{widetext} 
\begin{align}\label{eq:3D}
			\begin{aligned}
				\hspace{-0.226 cm}\Gamma_{\alpha\mu\nu\beta}^{(3D)} &=
				i^3 \dfrac{\langle \bar{s} s \rangle}{4 (2 \pi^2)^3}
				\Bigg\{ 
				2 \Big(-1 + \dfrac{i m_s (\yslash - \xslash)}{4}  \Big)
				\gamma_{\beta} \Big( i \dfrac{\yslash - \xslash}{(y-x)^{4}} + \dfrac{m_s}{2(y-x)^{2}}  \Big) \gamma_{\alpha} 
				E^P_{\mu\nu} (x,y)\\
				&+2 \Big( i \dfrac{\yslash - \xslash}{(y-x)^{4}} - \dfrac{m_s}{2(y-x)^{2}}  \Big)
				\gamma_{\beta} \Big(1 + \dfrac{i m_s (\yslash - \xslash)}{4} \Big) \gamma_{\alpha} 
				E^P_{\mu\nu} (x,y)\\
				&+
				2 E^P_{\mu\nu} (x,y)
				\gamma_{\beta} \Big(1 + \dfrac{i m_s (\yslash - \xslash)}{4} \Big)
				\gamma_{\alpha} 
				\Big( i \dfrac{\yslash - \xslash}{(y-x)^{4}} - \dfrac{m_s}{2(y-x)^{2}}  \Big)
				\\	
				&+
				2 E^P_{\mu\nu} (x,y)
				\gamma_{\beta} \Big( i \dfrac{\yslash - \xslash}{(y-x)^{4}} + \dfrac{m_s}{2(y-x)^{2}}  \Big)
				\gamma_{\alpha} 
				\Big(-1 + \dfrac{i m_s (\yslash - \xslash)}{4}\Big)
				\\	
				&+
				2 \Big(-1 + \dfrac{i m_s (\yslash - \xslash)}{4}\Big)
				Tr \Big[
				\gamma_{\beta} \Big( i \dfrac{\yslash - \xslash}{(y-x)^{4}} + \dfrac{m_s}{2(y-x)^{2}}  \Big) \gamma_{\alpha} 
				E^P_{\mu\nu} (x,y)
				\Big]
				\\
				&+
				2 \Big( i \dfrac{\yslash - \xslash}{(y-x)^{4}} - \dfrac{m_s}{2(y-x)^{2}}  \Big)
				Tr \Big[
				\gamma_{\beta} \Big(1 + \dfrac{i m_s (\yslash - \xslash)}{4}  \Big) \gamma_{\alpha} 
				E^P_{\mu\nu} (x,y)
				\Big]
				\\
				&+
				E^P_{\mu\nu} (x,y)
				Tr \Big[
				\gamma_{\beta} \Big( i \dfrac{\yslash - \xslash}{(y-x)^{4}} + \dfrac{m_s}{2(y-x)^{2}}  \Big)
				\gamma_{\alpha} 
				\Big(-1 + \dfrac{i m_s (\yslash - \xslash)}{4}\Big)
				\Big]
				\\
				&+
				E^P_{\mu\nu} (x,y)
				Tr \Big[
				\gamma_{\beta} \Big(1 + \dfrac{i m_s (\yslash - \xslash)}{4}  \Big)
				\gamma_{\alpha} 
				\Big( i \dfrac{\yslash - \xslash}{(y-x)^{4}} - \dfrac{m_s}{2(y-x)^{2}}  \Big)
				\Big]
				\\
				&+
				2 \Big(-1 + \dfrac{i m_s (\yslash - \xslash)}{4} \Big)
				\gamma_{\beta} 
				F^P_{\mu\nu} (x,y)
				\gamma_{\alpha} 
				\Big( i \dfrac{\yslash - \xslash}{(y-x)^{4}} - \dfrac{m_s}{2(y-x)^{2}}  \Big)
				\\
				&+
				2 \Big( i \dfrac{\yslash - \xslash}{(y-x)^{4}} - \dfrac{m_s}{2(y-x)^{2}}  \Big)
				\gamma_{\beta} 
				F^P_{\mu\nu} (x,y)
				\gamma_{\alpha} 
				\Big( -1 + \dfrac{i m_s (\yslash - \xslash)}{4} \Big)
				\\
				&+	
				2 \Big( i \dfrac{\yslash - \xslash}{(y-x)^{4}} - \dfrac{m_s}{2(y-x)^{2}}  \Big)
				\gamma_{\beta} \Big( i \dfrac{\yslash - \xslash}{(y-x)^{4}} + \dfrac{m_s}{2(y-x)^{2}}  \Big) \gamma_{\alpha} 
				E^3_{\mu\nu} (x,y)\\
				&+
				2 E^3_{\mu\nu} (x,y)
				\gamma_{\beta} \Big( i \dfrac{\yslash - \xslash}{(y-x)^{4}} + \dfrac{m_s}{2(y-x)^{2}}  \Big)
				\gamma_{\alpha} 
				\Big( i \dfrac{\yslash - \xslash}{(y-x)^{4}} - \dfrac{m_s}{2(y-x)^{2}}  \Big)
				\\	
				&+
				2 \Big( i \dfrac{\yslash - \xslash}{(y-x)^{4}} - \dfrac{m_s}{2(y-x)^{2}}  \Big)
				Tr \Big[
				\gamma_{\beta} \Big( i \dfrac{\yslash - \xslash}{(y-x)^{4}} + \dfrac{m_s}{2(y-x)^{2}}  \Big) \gamma_{\alpha} 
				E^3_{\mu\nu} (x,y)
				\Big]
				\\
				&+
				E^3_{\mu\nu} (x,y)
				Tr \Big[
				\gamma_{\beta} \Big( i \dfrac{\yslash - \xslash}{(y-x)^{4}} + \dfrac{m_s}{2(y-x)^{2}}  \Big)
				\gamma_{\alpha} 
				\Big( i \dfrac{\yslash - \xslash}{(y-x)^{4}} - \dfrac{m_s}{2(y-x)^{2}}  \Big)
				\Big]
				\\
				&+
				2 \Big( i \dfrac{\yslash - \xslash}{(y-x)^{4}} - \dfrac{m_s}{2(y-x)^{2}}  \Big)
				\gamma_{\beta} 
				F^3_{\mu\nu} (x,y)
				\gamma_{\alpha} 
				\Big( i \dfrac{\yslash - \xslash}{(y-x)^{4}} - \dfrac{m_s}{2(y-x)^{2}}  \Big)
				\Bigg\},
			\end{aligned}
		\end{align}
	\end{widetext}

	\begin{widetext} 
		\begin{align}\label{eq:4Dquark}
			\begin{aligned}
				\Gamma_{\alpha\mu\nu\beta}^{(4D,q)} &=
				\dfrac{i^5 g_s^2 \langle G^2 \rangle}{24 (8 \pi^2)^4} 
				\Bigg\{ \dfrac{1}{(y-x)^{4}} \Bigg(
				2 \Big( (\yslash - \xslash)\sigma^{\lambda\delta}
				+ \sigma^{\lambda\delta} (\yslash - \xslash) \Big)
				\gamma_{\beta} 
				\Big( (\yslash - \xslash)\sigma_{\lambda\delta}
				+ \sigma_{\lambda\delta} (\yslash - \xslash) \Big)
				\gamma_{\alpha} E^P_{\mu\nu} (x,y)
				\\
				&+ 2 E^P_{\mu\nu} (x,y) \gamma_{\beta}
				\Big( (\yslash - \xslash)\sigma^{\lambda\delta}
				+ \sigma^{\lambda\delta} (\yslash - \xslash) \Big)
				\gamma_{\alpha}  
				\Big( (\yslash - \xslash)\sigma_{\lambda\delta}
				+ \sigma_{\lambda\delta} (\yslash - \xslash) \Big)
				\\
				&+ 2 \Big( (\yslash - \xslash)\sigma^{\lambda\delta}
				+ \sigma^{\lambda\delta} (\yslash - \xslash) \Big)
				Tr \Big[ \gamma_{\beta} 
				\Big( (\yslash - \xslash)\sigma_{\lambda\delta}
				+ \sigma_{\lambda\delta} (\yslash - \xslash) \Big)
				\gamma_{\alpha} E^P_{\mu\nu} (x,y) \Big]
				\\ 
				&+ E^P_{\mu\nu} (x,y) Tr \Big[ \gamma_{\beta} 
				\Big( (\yslash - \xslash)\sigma^{\lambda\delta}
				+ \sigma^{\lambda\delta} (\yslash - \xslash) \Big)
				\gamma_{\alpha}  
				\Big( (\yslash - \xslash)\sigma_{\lambda\delta}
				+ \sigma_{\lambda\delta} (\yslash - \xslash) \Big) \Big]
				\\
				&- 2 \Big( (\yslash - \xslash)\sigma^{\lambda\delta}
				+ \sigma^{\lambda\delta} (\yslash - \xslash) \Big)
				\gamma_{\beta} F^P_{\mu\nu} (x,y) \gamma_{\alpha} 
				\Big( (\yslash - \xslash)\sigma_{\lambda\delta}
				+ \sigma_{\lambda\delta} (\yslash - \xslash) \Big)  
				\Bigg)
				\\
				&+ \Big( i \dfrac{\yslash - \xslash}{(y-x)^{4}} - \dfrac{m_s}{2(y-x)^{2}}  \Big)
				\gamma_{\beta} \Big( i \dfrac{\yslash - \xslash}{(y-x)^{4}} + \dfrac{m_s}{2(y-x)^{2}}  \Big) \gamma_{\alpha} E^{G^2}_{\mu\nu} (x,y)
				\\
				&+
				E^{G^2}_{\mu\nu} (x,y)
				\gamma_{\beta} \Big( i \dfrac{\yslash - \xslash}{(y-x)^{4}} + \dfrac{m_s}{2(y-x)^{2}}  \Big)
				\gamma_{\alpha} 
				\Big( i \dfrac{\yslash - \xslash}{(y-x)^{4}} - \dfrac{m_s}{2(y-x)^{2}}  \Big)
				\\	
				&+
				\Big( i \dfrac{\yslash - \xslash}{(y-x)^{4}} - \dfrac{m_s}{2(y-x)^{2}}  \Big)
				Tr \Big[
				\gamma_{\beta} \Big( i \dfrac{\yslash - \xslash}{(y-x)^{4}} + \dfrac{m_s}{2(y-x)^{2}}  \Big) \gamma_{\alpha} 
				E^{G^2}_{\mu\nu} (x,y)
				\Big]
				\\
				&+
				\dfrac{1}{2} E^{G^2}_{\mu\nu} (x,y)
				Tr \Big[
				\gamma_{\beta} \Big( i \dfrac{\yslash - \xslash}{(y-x)^{4}} + \dfrac{m_s}{2(y-x)^{2}}  \Big)
				\gamma_{\alpha} 
				\Big( i \dfrac{\yslash - \xslash}{(y-x)^{4}} - \dfrac{m_s}{2(y-x)^{2}}  \Big)
				\Big]
				\\
				&+
				\Big( i \dfrac{\yslash - \xslash}{(y-x)^{4}} - \dfrac{m_s}{2(y-x)^{2}}  \Big)
				\gamma_{\beta} 
				F^{G^2}_{\mu\nu} (x,y)
				\gamma_{\alpha} 
				\Big( i \dfrac{\yslash - \xslash}{(y-x)^{4}} - \dfrac{m_s}{2(y-x)^{2}}  \Big)
				\\
				&+\dfrac{1}{(y-x)^{2}} \Bigg(
				2 \Big( i \dfrac{\yslash - \xslash}{(y-x)^{4}} - \dfrac{m_s}{2(y-x)^{2}}  \Big)
				\gamma_{\beta} \Big( (\yslash - \xslash)\sigma^{\lambda\delta}
				+ \sigma^{\lambda\delta} (\yslash - \xslash) \Big) \gamma_{\alpha} 
				E^{PG}_{\mu\nu} (x,y)\\
				&- 2 \Big( (\yslash - \xslash)\sigma^{\lambda\delta}
				+ \sigma^{\lambda\delta} (\yslash - \xslash) \Big)
				\gamma_{\beta} \Big( i \dfrac{\yslash - \xslash}{(y-x)^{4}} + \dfrac{m_s}{2(y-x)^{2}}  \Big) \gamma_{\alpha} 
				E^{PG}_{\mu\nu} (x,y)\\
				&+
				2 E^{PG}_{\mu\nu} (x,y)
				\gamma_{\beta} \Big( (\yslash - \xslash)\sigma^{\lambda\delta}
				+ \sigma^{\lambda\delta} (\yslash - \xslash) \Big)
				\gamma_{\alpha} 
				\Big( i \dfrac{\yslash - \xslash}{(y-x)^{4}} - \dfrac{m_s}{2(y-x)^{2}}  \Big)
				\\	
				&-
				2 E^{PG}_{\mu\nu} (x,y)
				\gamma_{\beta} \Big( i \dfrac{\yslash - \xslash}{(y-x)^{4}} + \dfrac{m_s}{2(y-x)^{2}}  \Big)
				\gamma_{\alpha} 
				\Big( (\yslash - \xslash)\sigma^{\lambda\delta}
				+ \sigma^{\lambda\delta} (\yslash - \xslash) \Big)
				\\	
				&+
				2 \Big( i \dfrac{\yslash - \xslash}{(y-x)^{4}} - \dfrac{m_s}{2(y-x)^{2}}  \Big)
				Tr \Big[
				\gamma_{\beta} \Big( (\yslash - \xslash)\sigma^{\lambda\delta}
				+ \sigma^{\lambda\delta} (\yslash - \xslash) \Big) \gamma_{\alpha} 
				E^{PG}_{\mu\nu} (x,y)
				\Big]
				\\
				&-
				2 \Big( (\yslash - \xslash)\sigma^{\lambda\delta}
				+ \sigma^{\lambda\delta} (\yslash - \xslash) \Big)
				Tr \Big[
				\gamma_{\beta} \Big( i \dfrac{\yslash - \xslash}{(y-x)^{4}} + \dfrac{m_s}{2(y-x)^{2}}  \Big) \gamma_{\alpha} 
				E^{PG}_{\mu\nu} (x,y)
				\Big]
				\\
				&+
				E^{PG}_{\mu\nu} (x,y)
				Tr \Big[
				\gamma_{\beta} \Big( (\yslash - \xslash)\sigma^{\lambda\delta}
				+ \sigma^{\lambda\delta} (\yslash - \xslash) \Big)
				\gamma_{\alpha} 
				\Big( i \dfrac{\yslash - \xslash}{(y-x)^{4}} - \dfrac{m_s}{2(y-x)^{2}}  \Big)
				\Big]
				\\
				&-
				E^{PG}_{\mu\nu} (x,y)
				Tr \Big[
				\gamma_{\beta} \Big( i \dfrac{\yslash - \xslash}{(y-x)^{4}} + \dfrac{m_s}{2(y-x)^{2}}  \Big)
				\gamma_{\alpha} 
				\Big( (\yslash - \xslash)\sigma^{\lambda\delta}
				+ \sigma^{\lambda\delta} (\yslash - \xslash) \Big)
				\Big]
				\\
				&+
				2 \Big( i \dfrac{\yslash - \xslash}{(y-x)^{4}} - \dfrac{m_s}{2(y-x)^{2}}  \Big)
				\gamma_{\beta} 
				F^{PG}_{\mu\nu} (x,y)
				\gamma_{\alpha} 
				\Big( (\yslash - \xslash)\sigma^{\lambda\delta}
				+ \sigma^{\lambda\delta} (\yslash - \xslash) \Big)
				\\
				&+
				2 \Big( (\yslash - \xslash)\sigma^{\lambda\delta}
				+ \sigma^{\lambda\delta} (\yslash - \xslash) \Big)
				\gamma_{\beta} 
				F^{PG}_{\mu\nu} (x,y)
				\gamma_{\alpha} 
				\Big( i \dfrac{\yslash - \xslash}{(y-x)^{4}} - \dfrac{m_s}{2(y-x)^{2}}  \Big)
				\Bigg)
				\Bigg\}, 
			\end{aligned}
		\end{align}
	\end{widetext}

	\begin{widetext} 
		\begin{align}\label{eq:4Dgluon}
			\begin{aligned}
				\Gamma_{\alpha\mu\nu\beta}^{(4D,g)} &=
				\dfrac{6 i^2}{(2 \pi^2)^3} \langle G^2 \rangle g_{\mu\nu}
				\Big\{ 2 \Big( i \dfrac{\yslash - \xslash}{(y-x)^{4}} - \dfrac{m_s}{2(y-x)^{2}}  \Big)
				\gamma_{\beta} \Big( i \dfrac{\yslash - \xslash}{(y-x)^{4}} + \dfrac{m_s}{2(y-x)^{2}}  \Big) \gamma_{\alpha} \Big( i \dfrac{\yslash - \xslash}{(y-x)^{4}} - \dfrac{m_s}{2(y-x)^{2}}  \Big) \\
				&+ \Big( i \dfrac{\yslash - \xslash}{(y-x)^{4}} - \dfrac{m_s}{2(y-x)^{2}}  \Big)
				Tr \Big[
				\gamma_{\beta} \Big( i \dfrac{\yslash - \xslash}{(y-x)^{4}} + \dfrac{m_s}{2(y-x)^{2}}  \Big) \gamma_{\alpha} \Big( i \dfrac{\yslash - \xslash}{(y-x)^{4}} - \dfrac{m_s}{2(y-x)^{2}}  \Big)
				\Big] \Big\},
			\end{aligned}
		\end{align}
	\end{widetext}

\begin{widetext} 
\begin{align}\label{eq:5D}
\begin{aligned}
\Gamma_{\alpha\mu\nu\beta}^{(5D)} &=
				i^3 \dfrac{m_0^2\langle \bar{s} s \rangle}{(8 \pi^2)^3}
				\Bigg\{ 
				2 (y-x)^{2}\Big(-1 + \dfrac{i m_s  (\yslash - \xslash)}{6}  \Big)
				\gamma_{\beta} \Big( i \dfrac{\yslash - \xslash}{(y-x)^{4}} + \dfrac{m_s}{2(y-x)^{2}}  \Big) \gamma_{\alpha} 
				E^P_{\mu\nu} (x,y)\\
				&+2 (y-x)^{2}\Big( i \dfrac{\yslash - \xslash}{(y-x)^{4}} - \dfrac{m_s}{2(y-x)^{2}}  \Big)
				\gamma_{\beta} \Big(1 + \dfrac{i m_s (\yslash - \xslash)}{6} \Big) \gamma_{\alpha} 
				E^P_{\mu\nu} (x,y)\\
				&+
				2 (y-x)^{2} E^P_{\mu\nu} (x,y)
				\gamma_{\beta} \Big(1 + \dfrac{i m_s (\yslash - \xslash)}{6} \Big)
				\gamma_{\alpha} 
				\Big( i \dfrac{\yslash - \xslash}{(y-x)^{4}} - \dfrac{m_s}{2(y-x)^{2}}  \Big)
				\\	
				&+
				2 (y-x)^{2} E^P_{\mu\nu} (x,y)
				\gamma_{\beta} \Big( i \dfrac{\yslash - \xslash}{(y-x)^{4}} + \dfrac{m_s}{2(y-x)^{2}}  \Big)
				\gamma_{\alpha} 
				\Big(-1 + \dfrac{i m_s (\yslash - \xslash)}{6}\Big)
				\\	
				&+
				2 (y-x)^{2} \Big(-1 + \dfrac{i m_s (\yslash - \xslash)}{6}\Big)
				Tr \Big[
				\gamma_{\beta} \Big( i \dfrac{\yslash - \xslash}{(y-x)^{4}} + \dfrac{m_s}{2(y-x)^{2}}  \Big) \gamma_{\alpha} 
				E^P_{\mu\nu} (x,y)
				\Big]
				\\
				&+
				2 (y-x)^{2} \Big( i \dfrac{\yslash - \xslash}{(y-x)^{4}} - \dfrac{m_s}{2(y-x)^{2}}  \Big)
				Tr \Big[
				\gamma_{\beta} \Big(1 + \dfrac{i m_s (\yslash - \xslash)}{6}  \Big) \gamma_{\alpha} 
				E^P_{\mu\nu} (x,y)
				\Big]
				\\
				&+
				(y-x)^{2} E^P_{\mu\nu} (x,y)
				Tr \Big[
				\gamma_{\beta} \Big( i \dfrac{\yslash - \xslash}{(y-x)^{4}} + \dfrac{m_s}{2(y-x)^{2}}  \Big)
				\gamma_{\alpha} 
				\Big(-1 + \dfrac{i m_s (\yslash - \xslash)}{6}\Big)
				\Big]
				\\
				&+
				(y-x)^{2} E^P_{\mu\nu} (x,y)
				Tr \Big[
				\gamma_{\beta} \Big(1 + \dfrac{i m_s (\yslash - \xslash)}{6}  \Big)
				\gamma_{\alpha} 
				\Big( i \dfrac{\yslash - \xslash}{(y-x)^{4}} - \dfrac{m_s}{2(y-x)^{2}}  \Big)
				\Big]
				\\
				&+
				2 (y-x)^{2} \Big(-1 + \dfrac{i m_s (\yslash - \xslash)}{6} \Big)
				\gamma_{\beta} 
				F^P_{\mu\nu} (x,y)
				\gamma_{\alpha} 
				\Big( i \dfrac{\yslash - \xslash}{(y-x)^{4}} - \dfrac{m_s}{2(y-x)^{2}}  \Big)
				\\
				&+
				2 (y-x)^{2} \Big( i \dfrac{\yslash - \xslash}{(y-x)^{4}} - \dfrac{m_s}{2(y-x)^{2}}  \Big)
				\gamma_{\beta} 
				F^P_{\mu\nu} (x,y)
				\gamma_{\alpha} 
				\Big( -1 + \dfrac{i m_s (\yslash - \xslash)}{6} \Big)
				\\
				&+	
				2 \Big( i \dfrac{\yslash - \xslash}{(y-x)^{4}} - \dfrac{m_s}{2(y-x)^{2}}  \Big)
				\gamma_{\beta} \Big( i \dfrac{\yslash - \xslash}{(y-x)^{4}} + \dfrac{m_s}{2(y-x)^{2}}  \Big) \gamma_{\alpha} 
				E^5_{\mu\nu} (x,y)\\
				&+
				2 E^5_{\mu\nu} (x,y)
				\gamma_{\beta} \Big( i \dfrac{\yslash - \xslash}{(y-x)^{4}} + \dfrac{m_s}{2(y-x)^{2}}  \Big)
				\gamma_{\alpha} 
				\Big( i \dfrac{\yslash - \xslash}{(y-x)^{4}} - \dfrac{m_s}{2(y-x)^{2}}  \Big)
				\\	
				&+
				2 \Big( i \dfrac{\yslash - \xslash}{(y-x)^{4}} - \dfrac{m_s}{2(y-x)^{2}}  \Big)
				Tr \Big[
				\gamma_{\beta} \Big( i \dfrac{\yslash - \xslash}{(y-x)^{4}} + \dfrac{m_s}{2(y-x)^{2}}  \Big) \gamma_{\alpha} 
				E^5_{\mu\nu} (x,y)
				\Big]
				\\
				&+
				E^5_{\mu\nu} (x,y)
				Tr \Big[
				\gamma_{\beta} \Big( i \dfrac{\yslash - \xslash}{(y-x)^{4}} + \dfrac{m_s}{2(y-x)^{2}}  \Big)
				\gamma_{\alpha} 
				\Big( i \dfrac{\yslash - \xslash}{(y-x)^{4}} - \dfrac{m_s}{2(y-x)^{2}}  \Big)
				\Big]
				\\
				&+
				2 \Big( i \dfrac{\yslash - \xslash}{(y-x)^{4}} - \dfrac{m_s}{2(y-x)^{2}}  \Big)
				\gamma_{\beta} 
				F^5_{\mu\nu} (x,y)
				\gamma_{\alpha} 
				\Big( i \dfrac{\yslash - \xslash}{(y-x)^{4}} - \dfrac{m_s}{2(y-x)^{2}}  \Big)
				\Bigg\},
			\end{aligned}
		\end{align}
	\end{widetext}
	where,
	\begin{widetext} 
		\begin{align}\label{eq:A}
			\begin{aligned}
				&A^P_{\mu\nu} (x,y) = 
				\Big(i \dfrac{\yslash}{y^4} - \dfrac{m_s}{2 y^2} \Big)
				\Big[
				m_s g_{\mu\nu} \dfrac{\xslash}{x^4}
				+\gamma_{\nu}
				\Big(\dfrac{i \gamma_{\mu}}{x^4} - \dfrac{4 i \xslash x_{\mu}}{x^6} - m_s 
				\dfrac{x_{\mu}}{x^4} \Big)
				\Big] \\
				&\qquad\qquad+ 
				\Big[
				m_s g_{\mu\nu} \dfrac{\yslash}{y^4}
				-
				\Big(\dfrac{i \gamma_{\mu}}{y^4} - \dfrac{4 i \yslash y_{\mu}}{y^6} + m_s
				\dfrac{y_{\mu}}{y^4} \Big) \gamma_{\nu} 
				\Big]
				\Big(i \dfrac{\xslash}{x^4} + \dfrac{m_s}{2 x^2} \Big), 
				\\
				&A^3_{\mu\nu} (x,y) = 
				i m_s \Big(i \dfrac{\yslash}{y^4} - \dfrac{m_s}{2 y^2} \Big)
				\Big[
				\dfrac{\gamma_{\nu} \gamma_{\mu}}{4} - g_{\mu\nu} 
				\Big] 
				-
				\Big(1 - \dfrac{i m_s}{4} \yslash \Big)
				\Big[
				m_s g_{\mu\nu} \dfrac{\xslash}{x^4}
				+\gamma_{\nu}
				\Big(\dfrac{i \gamma_{\mu}}{x^4} - \dfrac{4 i \xslash x_{\mu}}{x^6} - m_s 
				\dfrac{x_{\mu}}{x^4} \Big)
				\Big]
				\\
				&\qquad\qquad 
				+
				i m_s 
				\Big[
				g_{\mu\nu} - \dfrac{\gamma_{\mu} \gamma_{\nu}}{4} 
				\Big]
				\Big(i \dfrac{\xslash}{x^4} + \dfrac{m_s}{2 x^2} \Big)
				+ 
				\Big[
				m_s g_{\mu\nu} \dfrac{\yslash}{y^4}
				-
				\Big(\dfrac{i \gamma_{\mu}}{y^4} - \dfrac{4 i \yslash y_{\mu}}{y^6} + m_s
				\dfrac{y_{\mu}}{y^4} \Big) \gamma_{\nu} 
				\Big]
				\Big(1 + \dfrac{i m_s}{4} \xslash \Big),
				\\
				&A^{G^2}_{\mu\nu} (x,y)	=	\Big(
				\dfrac{\yslash \sigma^{\lambda\delta} + \sigma^{\lambda\delta}\yslash}{y^2} \Big)
				\Big[\gamma_{\nu}  \bigg(\Big(
				\dfrac{\gamma_{\mu}}{x^2} - \dfrac{2 \xslash x_{\mu}}{x^4}
				\Big) \sigma_{\lambda\delta}
				+ \sigma_{\lambda\delta}
				\Big( \dfrac{\gamma_{\mu}}{x^2} - \dfrac{2 \xslash x_{\mu}}{x^4}
				\Big) \bigg)
				-\dfrac{2}{x^2} g_{\mu\nu} \Big( \sigma_{\lambda\delta} - \dfrac{\xslash \sigma_{\lambda\delta} \xslash}{x^2} \Big)
				\Big] 
				\\
				&\qquad\qquad - \Big[ \bigg( \Big(
				\dfrac{\gamma_{\mu}}{y^2} - \dfrac{2 \yslash y_{\mu}}{y^4}
				\Big) \sigma^{\lambda\delta}
				+ \sigma^{\lambda\delta}
				\Big( \dfrac{\gamma_{\mu}}{y^2} - \dfrac{2 \yslash y_{\mu}}{y^4} \Big)
				\bigg)\gamma_{\nu}
				-\dfrac{2}{y^2} g_{\mu\nu} \Big( \sigma^{\lambda\delta} - \dfrac{\yslash \sigma^{\lambda\delta} \yslash}{y^2} \Big)
				\Big]
				\Big(
				\dfrac{\xslash \sigma_{\lambda\delta} + \sigma_{\lambda\delta}\xslash}{x^2} \Big),
				\\
				&A^{PG}_{\mu\nu} (x,y)	=
				\Big(i \dfrac{\yslash}{y^4} - \dfrac{m_s}{2 y^2} \Big)
				\Big[\gamma_{\nu}  \bigg(\Big(
				\dfrac{\gamma_{\mu}}{x^2} - \dfrac{2 \xslash x_{\mu}}{x^4}
				\Big) \sigma_{\lambda\delta}
				+ \sigma_{\lambda\delta}
				\Big( \dfrac{\gamma_{\mu}}{x^2} - \dfrac{2 \xslash x_{\mu}}{x^4}
				\Big) \bigg)
				-\dfrac{2}{x^2} g_{\mu\nu} \Big( \sigma_{\lambda\delta} - \dfrac{\xslash \sigma_{\lambda\delta} \xslash}{x^2} \Big)
				\Big] 
				\\
				&\qquad\qquad + 
				\Big[
				m_s g_{\mu\nu} \dfrac{\yslash}{y^4}
				- \Big(\dfrac{i \gamma_{\mu}}{y^4} - \dfrac{4 i \yslash y_{\mu}}{y^6} + m_s
				\dfrac{y_{\mu}}{y^4} \Big) \gamma_{\nu} 
				\Big]
				\Big(
				\dfrac{\xslash \sigma_{\lambda\delta} + \sigma_{\lambda\delta}\xslash}{x^2} \Big)
				\\
				&\qquad\qquad +
				\Big(
				\dfrac{\yslash \sigma_{\lambda\delta} + \sigma_{\lambda\delta}\yslash}{y^2} \Big)
				\Big[
				m_s g_{\mu\nu} \dfrac{\xslash}{x^4}
				+ \gamma_{\nu}
				\Big(\dfrac{i \gamma_{\mu}}{x^4} - \dfrac{4 i \xslash x_{\mu}}{x^6} - m_s 
				\dfrac{x_{\mu}}{x^4} \Big)
				\Big] 
				\\
				&\qquad\qquad - \Big[ \bigg( \Big(
				\dfrac{\gamma_{\mu}}{y^2} - \dfrac{2 \yslash y_{\mu}}{y^4}
				\Big) \sigma_{\lambda\delta}
				+ \sigma_{\lambda\delta}
				\Big( \dfrac{\gamma_{\mu}}{y^2} - \dfrac{2 \yslash y_{\mu}}{y^4} \Big)
				\bigg)\gamma_{\nu}
				-\dfrac{2}{y^2} g_{\mu\nu} \Big( \sigma_{\lambda\delta} - \dfrac{\yslash \sigma_{\lambda\delta} \yslash}{y^2} \Big)
				\Big]
				\Big(i \dfrac{\xslash}{x^4} + \dfrac{m_s}{2 x^2} \Big), 
				\\
				&A^5_{\mu\nu} (x,y) = 
				\Big(i \dfrac{\yslash}{y^4} - \dfrac{m_s}{2 y^2} \Big)
				\Big[
				\gamma_{\nu} \Big(2 x_{\mu} \big[1 + \dfrac{i m_s}{6} \xslash \big] 
				+ \dfrac{i m_s}{6} x^2 \gamma_{\mu} \Big)
				- g_{\mu\nu} \Big(2 \xslash + i m_s x^2 \Big)
				\Big] 
				\\
				&\qquad\qquad
				+
				\Big[
				\Big(2 y_{\mu} \big[1 - \dfrac{i m_s}{6} \yslash \big] 
				- \dfrac{i m_s}{6} y^2 \gamma_{\mu} \Big) \gamma_{\nu} 
				- g_{\mu\nu} \Big(2 \yslash - i m_s y^2 \Big)
				\Big] 
				\Big(i \dfrac{\xslash}{x^4} + \dfrac{m_s}{2 x^2} \Big)
				\\
				&\qquad\qquad
				-
				y^2 \Big(1 - \dfrac{i m_s}{6} \yslash \Big)
				\Big[
				m_s g_{\mu\nu} \dfrac{\xslash}{x^4}
				+\gamma_{\nu}
				\Big(\dfrac{i \gamma_{\mu}}{x^4} - \dfrac{4 i \xslash x_{\mu}}{x^6} - m_s 
				\dfrac{x_{\mu}}{x^4} \Big)
				\Big]
				\\
				&\qquad\qquad 
				+ 
				x^2 \Big[
				m_s g_{\mu\nu} \dfrac{\yslash}{y^4}
				-
				\Big(\dfrac{i \gamma_{\mu}}{y^4} - \dfrac{4 i \yslash y_{\mu}}{y^6} + m_s
				\dfrac{y_{\mu}}{y^4} \Big) \gamma_{\nu} 
				\Big]
				\Big(1 + \dfrac{i m_s}{6} \xslash \Big),
				\\
				&R^P (x,y) = 
				\Big(i \dfrac{\yslash}{y^4} - \dfrac{m_s}{2 y^2} \Big)
				\Big(i \dfrac{\xslash}{x^4} + \dfrac{m_s}{2 x^2} \Big), 
				\\
				&R^3 (x,y) = 
				\Big(i \dfrac{\yslash}{y^4} - \dfrac{m_s}{2 y^2} \Big)
				\Big(1 + \dfrac{i m_s}{4} \xslash \Big) -
				\Big(1 - \dfrac{i m_s}{4} \yslash \Big)
				\Big(i \dfrac{\xslash}{x^4} + \dfrac{m_s}{2 x^2} \Big), 
				\\
				&R^{G^2} (x,y)	=	\Big(
				\dfrac{\yslash \sigma^{\lambda\delta} + \sigma^{\lambda\delta}\yslash}{y^2} \Big)
				\Big(
				\dfrac{\xslash \sigma_{\lambda\delta} + \sigma_{\lambda\delta}\xslash}{x^2} \Big),
				\\
				&R^{PG} (x,y) = 
				\Big(i \dfrac{\yslash}{y^4} - \dfrac{m_s}{2 y^2} \Big)
				\Big(
				\dfrac{\xslash \sigma_{\lambda\delta} + \sigma_{\lambda\delta}\xslash}{x^2} \Big)
				+ \Big(
				\dfrac{\yslash \sigma_{\lambda\delta} + \sigma_{\lambda\delta}\yslash}{y^2} \Big)
				\Big(i \dfrac{\xslash}{x^4} + \dfrac{m_s}{2 x^2} \Big),
				\\
				&R^5 (x,y) = 
				x^2 \Big(i \dfrac{\yslash}{y^4} - \dfrac{m_s}{2 y^2} \Big)
				\Big(1 + \dfrac{i m_s}{6} \xslash \Big) -
				y^2 \Big(1 - \dfrac{i m_s}{6} \yslash \Big)
				\Big(i \dfrac{\xslash}{x^4} + \dfrac{m_s}{2 x^2} \Big), 
				\\
				&B_{\mu\nu} (x,y) = A_{\mu\nu} (y,x), \\ 
                &Z (x,y) = R (y,x), \\
				&E_{\mu\nu} (x,y) = A_{\mu\nu} (x,y) + 2 i m_s g_{\mu\nu} R (x,y), \\
				&F_{\mu\nu} (x,y) = B_{\mu\nu} (x,y) - 2 i m_s g_{\mu\nu} Z (x,y). 
\end{aligned}
\end{align}
\end{widetext}

	
\end{document}